\documentclass[12pt,preprint]{aastex}

\shorttitle{Six SMC fields}
\shortauthors{Sabbi et al.}

\begin{document}


\title{Star Formation History of the SMC: six HST/ACS fields\footnote{Based on observations with the NASA/ESA Hubble Space
Telescope, obtained at the Space Telescope Science Institute, which is
operated by AURA Inc., under NASA contract NAS 5-26555. These observations are associated with program GO-10396.}}


\author{E. Sabbi\altaffilmark{1}, J.S. Gallagher\altaffilmark{2}, M. Tosi\altaffilmark{3}, J. Anderson\altaffilmark{1}, A. Nota\altaffilmark{1,4}, E.K. Grebel\altaffilmark{5}, M. Cignoni\altaffilmark{2}, A.A. Cole\altaffilmark{6}, G.S. Da Costa\altaffilmark{7}, D. Harbeck\altaffilmark{3}, K. Glatt\altaffilmark{4,8}, \& M. Marconi \altaffilmark{9}} 

\email{sabbi@stsci.edu}


\altaffiltext{1}{STScI, San Martin Drive, Baltimore, MD, 21218, USA}
\altaffiltext{2}{University of Wisconsin, Madison, WI, USA}
\altaffiltext{3}{INAF-Osservatorio Astronomico di Bologna, Bologna, Italy}
\altaffiltext{4}{European Space Agency, Research and Scientific Support Department, Baltimore, USA}
\altaffiltext{5}{Astronomisches Rechen-Institut, Zentrum f\"{u}r Astronomie der Universit\"{a}t Heidelberg, Heidelberg, Germany}
\altaffiltext{6}{School of Mathematics \& Physics, Tasmania, Australia}
\altaffiltext{7}{Research School of Astronomy \& Astrophysics, The Australian National University, Weston, Australia}
\altaffiltext{8}{Astronomical Institute, Department of Physics and Astronomy,
University of Basel, Binningen, Switzerland}
\altaffiltext{9}{INAF-Osservatorio Astronomico di Capodimonte, Napoli, Italy}


\begin{abstract}

We observed six fields of the Small Magellanic Cloud (SMC) with the Advanced Camera for Survey on
board the Hubble Space Telescope in the F555W and F814W filters. These fields sample regions
characterized by very different star and gas densities, and, possibly, by different evolutionary
histories. We find that the SMC was already forming stars $\sim 12$ Gyr ago, even if the lack of a
clear horizontal branch suggests that in the first few billion years the star formation activity was
low. Within the uncertainties of our two-band photometry, we find evidence of a radial variation in
chemical enrichment, with the SMC outskirts characterized by lower metallicity than the central
zones. From our CMDs we also infer that the SMC formed stars over a long interval of time until
$\sim 2-3$ Gyr ago. After a period of modest activity, star formation increased again in the recent
past, especially in the bar and the wing of the SMC, where we see an enhancement in the
star-formation activity starting from $\sim 500$ Myr ago. The inhomogeneous distribution of stars
younger than $\sim 100$ Myr indicates that recent star formation has mainly developed locally. 

\end{abstract}





\section{Introduction}

Dwarf galaxies are the most common galaxies in the Local Group (LG), and probably in the entire
universe.
As the closest late-type star-forming dwarf, the SMC is an ideal laboratory for a detailed study of
these very common objects: its low dust and high gas content, as well as its low present-day
metallicity (Z=0.004), make it an excellent local analog to the majority of dwarf irregulars and
blue compact galaxies, whose metallicity distribution is peaked at this mean value. As a member of a
triple system, the SMC is also a natural benchmark to study interaction-driven modulations of the SF
activity. Furthermore its proximity \citep[60.6 kpc, corresponding to a distance modulus
$(m-M)_0=18.9$,][]{hilditch05} allows us to resolve into single stars its stellar populations well
below its oldest main sequence turn off (MSTO), and thus to study in detail its star formation
history (SFH) over a Hubble time by exploring ages, metallicities, and spatial distribution of its
stellar content.


The properties of old (age$\la 10$ Gyr) and intermediate (2$\la$ age$\la 10$ Gyr) stellar
populations can help us understand the formation and evolution of the LG. Deep photometric studies
of a number of dwarf galaxies have demonstrated that some of them started to form stars at the same
epoch: the oldest globular clusters in the Milky Way (MW), the Large Magellanic Cloud (LMC),
Sagittarius, and Fornax are coeval, and similar results have been obtained for the oldest stars in
the fields of the Milky Way, Fornax, Draco, Ursa Minor, Sculptor, Carina and Leo{\sc ii}
\citep{grebel04}. By contrast, the oldest (and only) globular cluster in the SMC is $\sim 2-3$ Gyr
younger than the oldest globulars in the Milky Way \citep{glatt08b}. 

While it is clear that the oldest surviving SMC cluster is younger than the oldest clusters in other
LG galaxies \citep[e.g.][]{glatt08a}, we still do not know whether the onset of SF in the SMC, and
possibly in other galaxies, was also delayed compared to the MW. To answer this question we need to
analyze the SMC field-star population. However, thus far only a few authors have investigated the SF
history (SFH) of field stars in the SMC. A first evidence of a stellar population older than 9-10
Gyr comes from the detection of RR-Lyrae stars \citep{graham75}. However the overall small number
($\sim 2\times 10^{-5}$ arcsec$^{-2}$) of RR-Lyrae found in the SMC bar \citep{soszynski02} suggests
that in the first 2-3 Gyr the star formation rate  in the SMC was low.  

\citet{harris04} have performed the most extensive study of the SFH of the SMC, covering $4\degr
\times 4.5\degr$ to a depth of $V\la 21$. 
They found that the SMC formed approximately 50\% of all its stars more than 8.4 Gyr ago, and only a
few stars between 3 and 8.4 Gyr ago. In the last 3 Gyr, the SMC formed stars at a constant rate,
with two bursts occurred approximately 2.5 and 0.4 Gyr ago. \citet{harris04} suggested that these
two events may have been triggered by the perigalactic passages of the SMC with the Milky Way. 

Other authors have focused on smaller regions, but with deeper and better photometry. The Wide Field Planetary Camera 2 (WFPC2) on board of the Hubble Space Telescope (HST) was used to analyzed the stellar content of the SMC wing \citep{mccumber05}, and in the vicinity of some SMC star clusters \citep[e.g.][]{chiosi07}. According to \citet{mccumber05} wing stars formed during a long episode of formation between 3.5 and 12 Gyr ago, followed by a period of quiescence with a very recent ($\sim 100$ Myr ago) pick up in SF activity. \citet{chiosi07} found periods of enhanced SF 0.3-0.4, 3-4 and 6 Gyr ago, and that SF in the SMC was likely quiescent at earlier epochs. 
\citet {dolphin01} presented the first analysis of the stellar content in the outskirts of the SMC using ground-based data. They found that stars in the outskirts of the SMC formed during a prolonged episode of SF between 2.5 and 15 Gyr ago, with a progressive chemical enrichment from [Fe/H]$=-$1.4 up to [Fe/H]$=-$0.7.

\citet{noel07} presented a deep ground-based study of 12 fields of the SMC. Due to the high crowding conditions, they avoided the densest regions, such as the bar. In all their fields the intermediate-age ($\sim 3-9$ Gyr) population is the dominant component, while the younger stars have a more asymmetric distribution, peaked in the wing. 
None of their fields showed a clear extended horizontal branch (HB), which would indicate a very old stellar population. From this they concluded that the SMC formed only a few stars at the beginning, and experienced an intense and prolonged period of SF between 1 and 6 Gyr ago. Finally, \citet{cignoni09} derived the SFH of the HST field around the young star-forming cluster NGC~602 in the SMC wing, where many pre-main sequence stars are also present, finding that the current SF activity is similar to that of Galactic OB associations, that the bulk of the old stars are 6-8 Gyr old, and with only few stars older than $\sim$8 Gyr.

In order to get a more accurate and complete evaluation of the SFH of the SMC from the earliest epochs up to the present day, and also to extend the analysis to the densest regions, we have started a long-term project (The Small Magellanic Cloud in Space and Time) aimed at getting deep and tight color magnitude diagrams (CMDs) of several selected SMC fields with HST, and in addition covering the whole galaxy down to several magnitudes below the oldest MSTO using the VLT Survey Telescope \citep[VST;][]{ripepi06}. The primary goals of this project are to combine deep and accurate photometry with high and intermediate-resolution spectra in order to map the SFH and abundance variations across the whole galaxy. These data will allow us to determine the age-metallicity relation (if there is one) of stars in different regions of the galaxy and to better constrain numerical models for the chemical evolution of the various SMC regions as well as for the galaxy as a whole. 
We will study the spatial distribution of the various types of variables and this will provide unique information on the space and time confinement of the formation of their parent stellar populations and on the 3D structure of the galaxy.
Finally the SFH will be derived from the color-magnitude diagrams using the synthetic CMD technique to study how, when and where the SF occurred in the SMC.
By combining the age-metallicity relation with the SFH through the galaxy we will be able to model the chemical evolution of the SMC with unprecedented accuracy for an external galaxy, reaching the same level of reliability that has been attained only for the solar neighborhood.

In this paper we present the stellar content of six fields observed with HST in the SMC bar (SFH1, SFH4, and SFH5), wing (SFH9 and SFH10) and the SMC outskirts (SFH8). All the regions have been observed with the Wide Field Channel (WFC) of the Advanced Camera for Surveys (ACS). To our knowledge this study represents the deepest and highest-resolution survey ever published for SMC fields. The SFH, as derived from the synthetic CMD technique, will be presented in a forthcoming paper (Cignoni et al., in preparation).

The paper is organized as follows: In section \S~\ref{obs} we present the data sets and discuss their reduction. Section \S~\ref{AT} presents artificial-star tests and photometric completeness. The fields are discussed in \S~\ref{6fields}, while a more detailed characterization of the stellar content of the various SMC regions (bar, wing and outskirts) is given in \S~\ref{stellar content}. Section \S~\ref{conclusions} summarizes our conclusions.

\section{Observations and data reduction}
\label{obs}

Six SMC fields were observed with the ACS/WFC between November 2005 and January 2006, as a part of a project (GO-10396; P.I. J. S. Gallagher, III) devoted to characterizing the properties of intermediate and old clusters \citep{glatt08a,glatt08b}, as well as to studying the field populations to infer the SFH of the SMC. The fields were selected to sample the stellar content in regions characterized by different stellar and gas density, and possibly different SFHs. The locations of the observed fields within the SMC are shown in Figure~\ref{f:targets}.

For each field we acquired 3-4 $\sim 500$ seconds long exposures both with the F555W ($\sim$ V) and F814W ($\sim$ I) filters, \citep[e.g. ][for a detailed description of ACS filters]{sirianni05}, and shorter exposures to recover the brighter stars that saturate in the longer exposures. The exposure times and dates of observations are listed in Table~\ref{t:obs}. The observations are 
The long exposures were acquired following a standard dither pattern that allowed us to cover the gap between the two chips of the ACS/WFC detector, to better sample the PSF and to better differentiate real sources from warm pixels or cosmic rays. 

WFC images cover an area of $200\arcsec \times 200\arcsec$ with a scale of $0.05\arcsec/$pixel. From the analysis of more than 50 eclipsing binaries, scattered over the whole galaxy, that therefore should be well representative of the mean distance value, \citet{hilditch05} derived a distance modulus $(m-M)_0=18.912\pm 0.035$ for the SMC. Corresponding to a distance of $\sim 60.6$ kpc, this implies that in each field we observed an area $\sim 50 \times 50$ pc$^2$. The data sets were processed according to the standard Space Telescope Science Institute ACS calibration pipeline (CALACS), which subtracts the bias level and applies the flat-field and overscan corrections. The pipeline-corrected ({\sc flt}) images are still in the raw-detector pixel frame and therefore retain several kinds of distortion. 

The photometric analysis was carried out directly on the {\sc flt} images. For each pointing and filter, positions and fluxes for all of the brighter stars (signal-to-noise ratio S/N$> 100$) were measured in all the {\sc flt} images, using the program img2xym\_WFC.09x10 \citep{anderson06}. The program uses a library of empirical PSFs, that take into account spatial variations in the WFC PSF, due to the telescope optics and the variable charge diffusion in the CCD \citep{krist03}. The PSF in each image can differ from the library PSF due to focus changes caused by spacecraft breathing. It is possible to take into account these differences by introducing for each exposure an appropriate spatially constant perturbation. This perturbation is obtained by fitting the library PSF to the brighter (S/N$> 100$) stars. In doing this we obtained a tailor-made PSF for each image. For each pointing we adopted a F555W deep exposure as the ``reference'' frame and we found a six parameter linear transformation from the distortion corrected frame of each {\sc FLT} image into this frame, using the position of the stars in common. 

For each of the {\sc FLT} images the position of all the stars was transformed in the reference frame and an average position and flux was computed for each filter. The root mean square residuals for the average positions of the stars were better than 0.01 pixel in each coordinate, and the photometric residuals were about 0.01 mag for the bright stars (S/N$> 100$). 

Using the reference list of stars for each field, we computed the final photometric zero points from both short and long exposures. Once we had created a photometric and astrometric mapping for each exposure, we used the program described in \citet{anderson08}, designed for the Globular Cluster Treasury Program (GO-10775. P.I. A. Sarajedini), to refine the photometry. This routine analyzes the reference frame with 25$\times 25$ pixel patch at a time in an automated way. For each patch, the program reads in the relevant portion of each exposure and searches all the exposures simultaneously in multiple passes. The brightest stars are found and subtracted first to allow the finding of the fainter stars.

Once all the stars have been found, the program fits all the exposures simultaneously for a single reference-frame position and a 2 filters (F555W and F814W in our case) flux. The process is then repeated for the next patch until the entire field is covered.

In addition to the simultaneous fit of fluxes, the program also measures each star in each individual FLT exposure where it could be found. After subtracting its neighbors, the routine determines a sky value in an annulus between 3 and 7 pixels from the center of the star for the fainter stars and between 4 and 8 pixels for the brighter stars, and then fits the PSF to the central $5 \times 5$ pixels. In this way for each field in each of the two filters we obtained between three and four independent estimates of the position and magnitude for each star in the long exposures, and 1 or 2 in the short one.

To remove as many spurious detections as possible from the final catalogs, we considered for each filter only the stars that were identified in at least three exposures with a positional error smaller than $<0.1$ pixel. With these selection criteria, the final catalog of the field SFH1 contains 29217 stars, 17384 stars were found in SFH4, 19769 stars in SFH5, 9180 stars in SFH10, and only 2659 stars in SFH9 and 1564 stars in SFH8. 

The calibration recipe by \citet{sirianni05} for all the filters available in ACS is derived for the drizzled images, thus this calibration cannot be directly applied to our photometric catalogs. We performed aperture photometry of the drizzled images using DAOPHOT in IRAF\footnote{IRAF is distributed by the National Optical Astronomy Observatory, which is operated by AURA, Inc., under cooperative agreement with the National Science Foundation} to obtain Vegamag calibrated magnitudes for both the F555W and F814W filters for many isolated stars in each of our fields, and used these stars to convert the instrumental magnitudes into the Vegamag system.

\subsection{Artificial-star tests}
\label{AT}

Artificial-star tests are a standard procedure to test the level of completeness and accuracy of a photometric analysis. The tests are performed by inserting stars with known position and flux into the dataset, and then repeating the photometric analysis as was used for the real stars. The difference between the input and output magnitudes of the recovered artificial stars provides an estimate of the photometric error and shows if, and how much, the data are affected by blending of nearby unresolved objects.

The routine used for the photometry analyzes the reference frame with 25$\times 25$ pixel patch at a time and, for each patch, the program reads the relevant portion of each exposure and searches all the exposures simultaneously in multiple passes. The same routine \citep{anderson08} allows us to add one artificial star in a patch and then it analyzes the data in exactly the same way described in \S~\ref{obs}. This approach avoids the concern that artificial stars may interact with each other. For each field we simulated more than 1,000,000 artificial stars in each of the F555W and F814W deep exposures. 

In our analysis we considered an artificial-star as recovered if its input and output fluxes agree to within 0.75 magnitudes. As in the real-stars photometric analysis, we also required that each star be found in at least three exposures with a positional error $<0.1$ pixels for filter.

\begin{figure}
\plotone{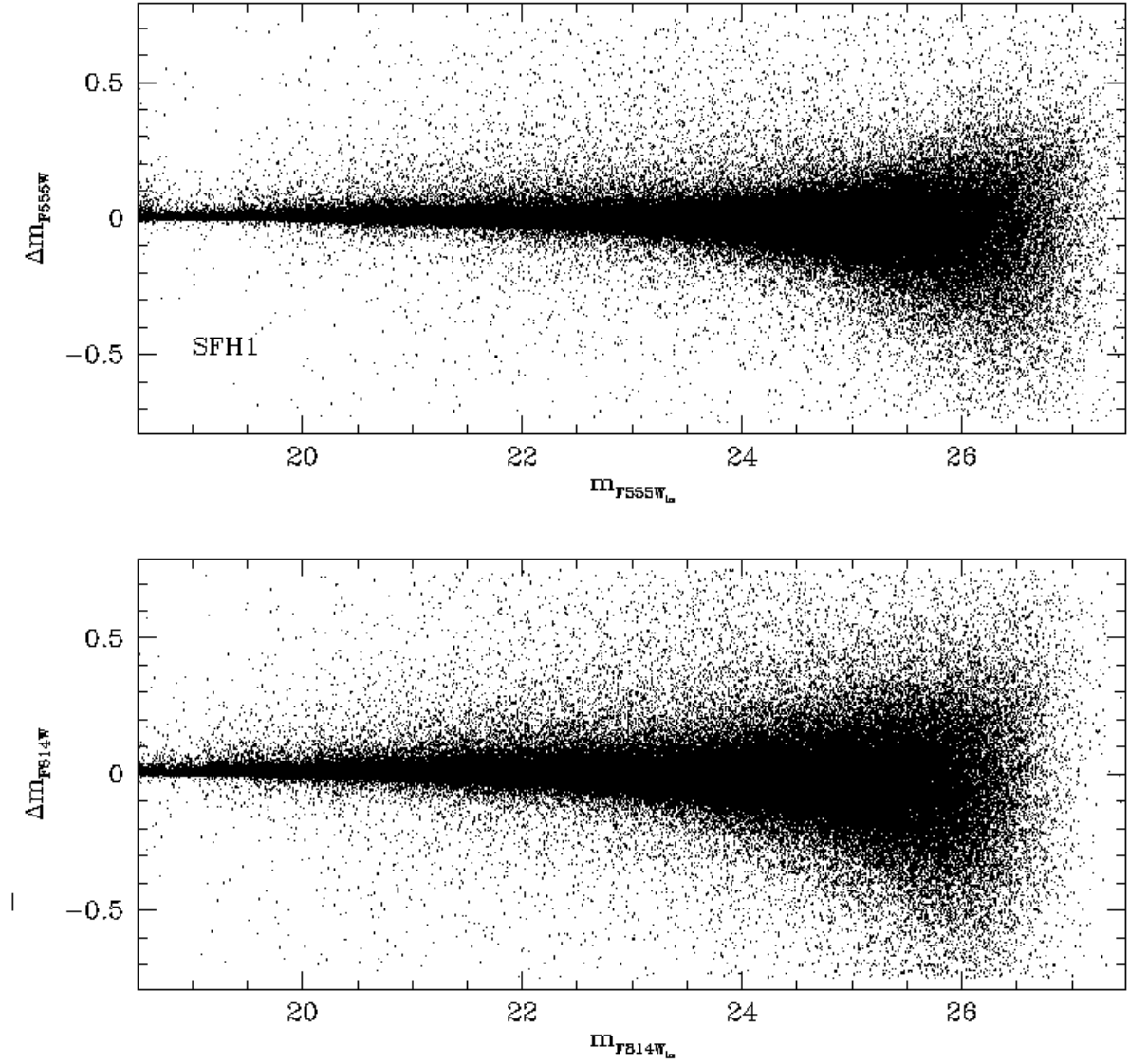}
\caption{\label{f:dmi} Diagrams of magnitude differences (input-output) vs. input magnitude from the artificial star experiments in the F555W and F814W filters for SFH1.}
\end{figure}

\begin{figure}
\plotone{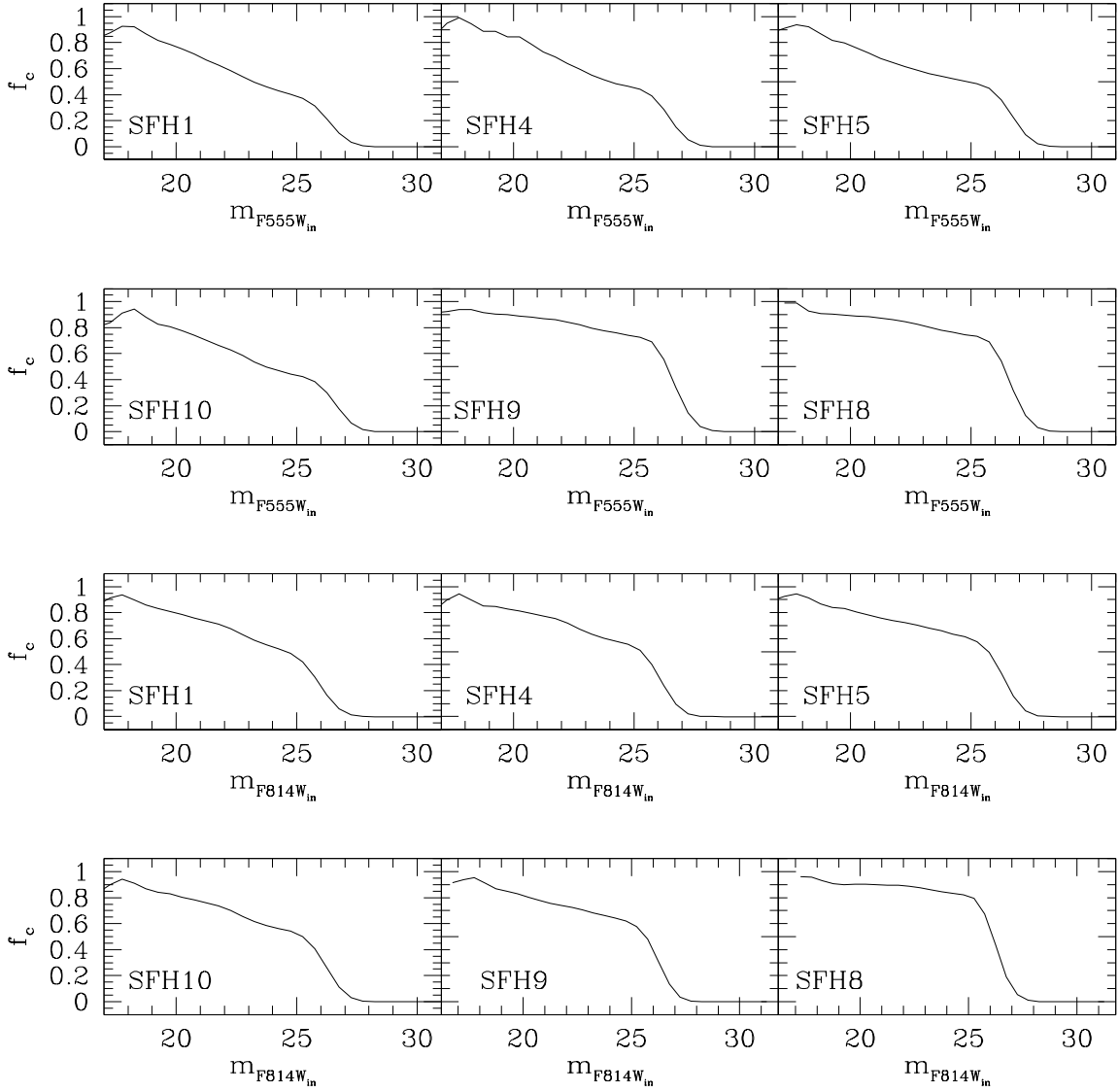}
\caption{\label{f:compl} Completeness fraction for each of the six observed SMC fields both in the F555W (six upper panels) and F814W (six lower panels) filters, as derived from the artificial star experiments.}
\end{figure}

Figure~\ref{f:dmi} shows the difference between input and output magnitudes of the recovered artificial-stars in both the filters for SFH1. The completeness fraction of each field is shown in Figure~\ref{f:compl}. Photometric errors, as derived from the difference between input and output magnitudes of the recovered artificial-stars, are plotted on the right side of the color-magnitude diagrams (CMDs) in Figure~\ref{f:cmds}.

\section{Properties of the six SMC fields}
\label{6fields}

In order to get a more accurate and complete evaluation of the SFH of the SMC from the earliest epochs up to the present day, and also to extend the analysis to the densest regions, we have observed six fields in the SMC with ACS/WFC. The fields were selected to sample regions characterized by different stellar and gas densities, such as the SMC bar, wing and the outskirts. Our survey was focused on the stellar content in the field of the SMC, therefore we avoided any known OB association, and star cluster. Because of the complex history of tidal interactions of the SMC with the LMC, and between the Magellanic Clouds and the MW it is possible that these regions have different histories of SF. 
 
\begin{figure}
\plotone{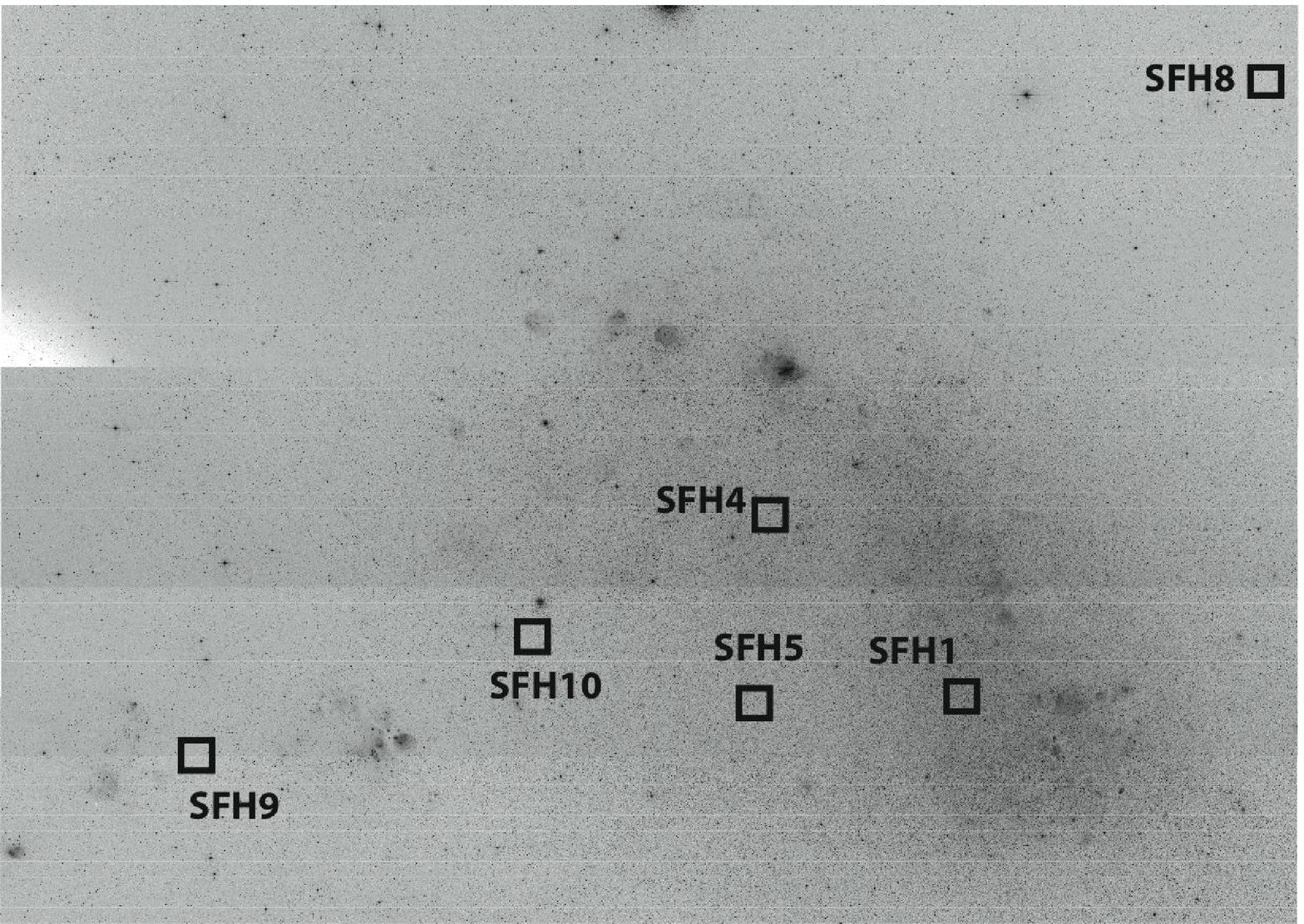}
\caption{\label{f:targets} Spatial distribution of the six observed fields (black squares), superimposed on the mosaic of images acquired by the Magellanic Cloud Emission Line Survey (MCELS). North is up, and East is left.}
\end{figure}

We selected three fields in the SMC bar (SFH1, SFH4, and SFH5 -- Fig.~\ref{f:targets}). SFH1 is the region closest ($\sim 24\arcmin$) to the optical center of the SMC. It is located to the SW of the SMC bar, where the stellar density, gas and dust content are highest. Ground-based photometry \citep{zaritsky00,harris04}, as well as several ionized nebulae \citep{bica08} indicate recent episodes of SF. The identification of young stellar objects (YSOs) associated with many of the ionized nebulae \citep{bolatto07} shows that SF is still high in the region.

SFH4 is located to the NE of the SMC center, $\sim 1\degr 52\arcmin$ from the optical center of the SMC, and $\sim 24\arcmin$ south of NGC~346, the most active star forming region in the SMC. Finally SFH5 is at the boundary between the SMC bar and the wing, $\sim 2\degr$ north of the optical center of the SMC.

In optical images, the SMC wing appears as a large cloud of faint stars, located to the east side of the SMC bar, toward the LMC, and is considered to be part of a tidal tail torn off the main body of the SMC by tidal interactions between the two Magellanic Clouds (MCs). We selected two regions to characterize the stellar populations in the wing: SFH10 in the center of the wing, and SFH9 closer to the Magellanic bridge connecting the two Clouds (Fig.~\ref{f:targets}).  

The properties of regions in the outskirts of the SMC are poorly constrained. Recently \citet{noel07b} detected intermediate age and old stars that belong to the SMC at $\sim 6.5$ kpc from the center of the SMC in the southern direction, suggesting that the SMC may be larger than previously thought. To investigate the properties of the stars in another part of the outskirts of the SMC we observed one field (SFH8) located at $\sim 3\degr 13\arcmin$ north-west from the SMC optical center (in the opposite direction with respect to the wing). Being far from the regions of recent SF, SFH8 will give us the best chance to isolate the oldest and most metal-poor stellar populations.  
 
\subsection{The Color-Magnitude Diagrams}
\label{cmds}

\begin{figure}
\plotone{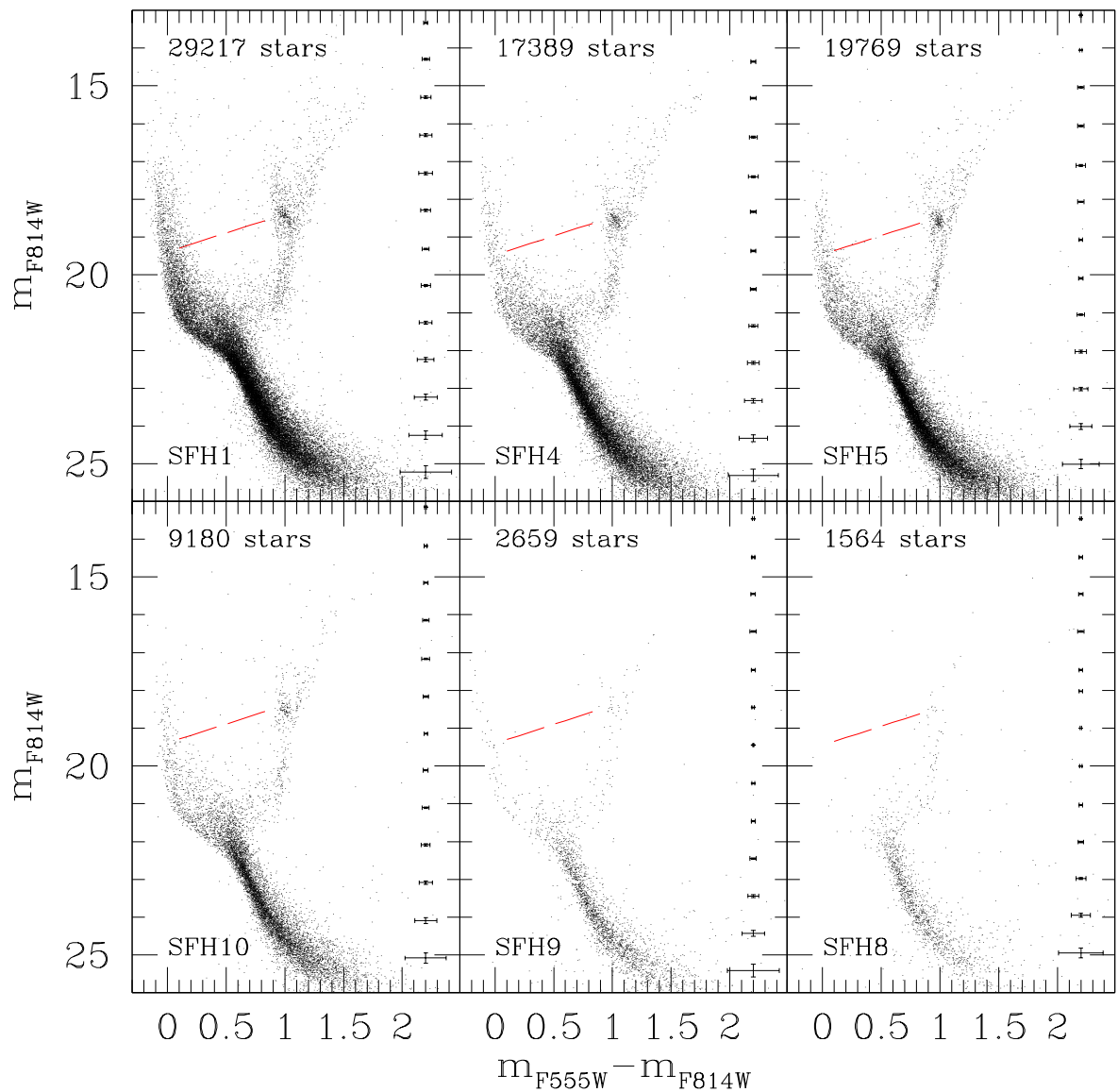}
\caption{\label{f:cmds} CMDs m$_{\rm F814W}$ vs. $m_{\rm F555W}-m_{\rm F814W}$ of the six SMC fields observed with the ACS/WFC. In each plot the red dashed line indicates the nominal location of the HB. Photometric errors, as derived from the artificial-star tests are shown on the right side of the CMDs.}
\end{figure}

Figure~\ref{f:cmds} shows the $m_{\rm F814W}$ versus $m_{\rm F555W}-m_{\rm F814W}$ CMDs obtained from the photometric analysis of the six fields. In all the regions our photometry reaches $\sim 3.5$ magnitudes below the faintest/oldest MSTO, that, assuming a distance modulus $(m-M)_0=18.9$ and a metallicity Z=0.001, according to Padua isochrones \citep{bertelli94} should be at $m_{\rm F814W}=22.48$. 

As in all resolved galaxies, there is a clear stellar density gradient from the center of the SMC to its periphery; the field containing the largest number of stars (29217; SFH1) is in the bar, and that with the lowest number (1564) in the SMC outskirts (SFH8). With the exception of SFH8, all the fields show the blue plume, typical of late-type dwarf galaxies, populated by high- and intermediate- mass stars in the main-sequence (MS) phase or at the blue edge of the blue loops (corresponding to the central He-burning phase). It is interesting to note the contrast between the two most external fields SFH8 and SFH9: in the latter, located in the outer wing, in spite of the low stellar density the blue plume is well populated by young stars.

An inspection by eye of the six CMDs reveals that stellar populations of very different ages are present in each area. 

\underline{Young stars}: With the exception of SFH8 (Fig.~\ref{f:cmds}), all the CMDs present a stellar population younger than $\sim 1-2$ Gyr, as indicated by the bright ($16.1 \la m_{\rm F814W} \la 21$) and blue ($-0.2 \la m_{\rm F555W} - m_{\rm F814W} \la 0.3$) well populated MS. This is also supported by the concentration of stars above the red clump (RC) in the magnitude range $16.0\la m_{\rm F814W} \la 18.0$ and color ranges $0.7\la m_{\rm F555W}-m_{\rm F814W}\la 1.05$, that corresponds to the lower end of the blue edge of the blue loop.
A careful inspection reveals differences from CMD to CMD, suggesting that recent (age$\la 100$ Myr) episodes of SF might have developed locally. 

\underline{Intermediate-age stars}: A stellar population older than $\sim 2$ Gyr is easily distinguishable in each CMD. Evolutionary phases associated with this population are the well defined lower MS, which extends from $m_{\rm F814W}\simeq 21.7$ down to $m_{\rm F814W}\simeq 25.7$, the broad subgiant branch (SGB), visible between $20.6 \la m_{\rm F814W} \la 21.6$ in the color range $0.45 \la m_{\rm F555W} - m_{\rm F814W} \la 0.95$, the red giant branch (RGB), well defined by the brightest stars at $m_{\rm F814W}\simeq 15.1 - 15.3$ and $m_{\rm F555w}-m_{\rm F814W} \simeq 1.4-1.5$, and the red clump (RC) at $m_{\rm F814W}\sim 18.5$. Asymptotic-giant-branch (AGB) stars seem to be present as well, but in most cases are difficult to distinguish from the RGB.

\underline{Old stars:} The dashed lines in Figure~\ref{f:cmds} indicate the nominal position of the HB. HB stars are unequivocal indicators of a stellar population older than $\sim 10$ Gyr. A few stars in SFH1, 4, and 5 have colors and magnitudes consistent with being HB stars, however their low number does not allow us to confirm the presence of HB stars in any of the analyzed regions. This may imply either that the SMC formed very few stars in the first 2-3 Gyr, or that even the oldest stars have a relatively high metallicity. Even so, many RR-Lyrae stars have been identified in the whole SMC \citep[e.g.][]{soszynski02}, confirming the presence of a metal-poor stellar population older than 10 Gyr. 
Future surveys devoted to identifying all the variable stars in the SMC \citep[i.e. with the ESO survey telescopes VISTA, in the IR, and VST, in the optical bands; see][respectively]{cioni08,ripepi06} will provide a complete census of the RR-Lyrae stars across the entire SMC, thus allowing us to fully trace the spatial distribution of the oldest populations. If some of the possible HB stars are eventually found to be RR-Lyraes, this will confirm the presence of an HB in these fields. 
In \S~\ref{stellar content} we will compare the properties of the MSTO and the evolved phases (SGB and RGB) to theoretical isochrones to investigate if a stellar population older than $\sim 10$ Gyr is present. 

\subsection{RCs and Distance Modulus}

At first glance, the stellar populations older than $\sim 2$ Gyr in the six fields appear fairly similar to each other, with no striking variations in the magnitudes of the SGBs, in the colors of the old MSTOs or in the magnitude and color of the RCs. However, a more careful analysis shows some interesting differences.

\begin{figure}
\plotone{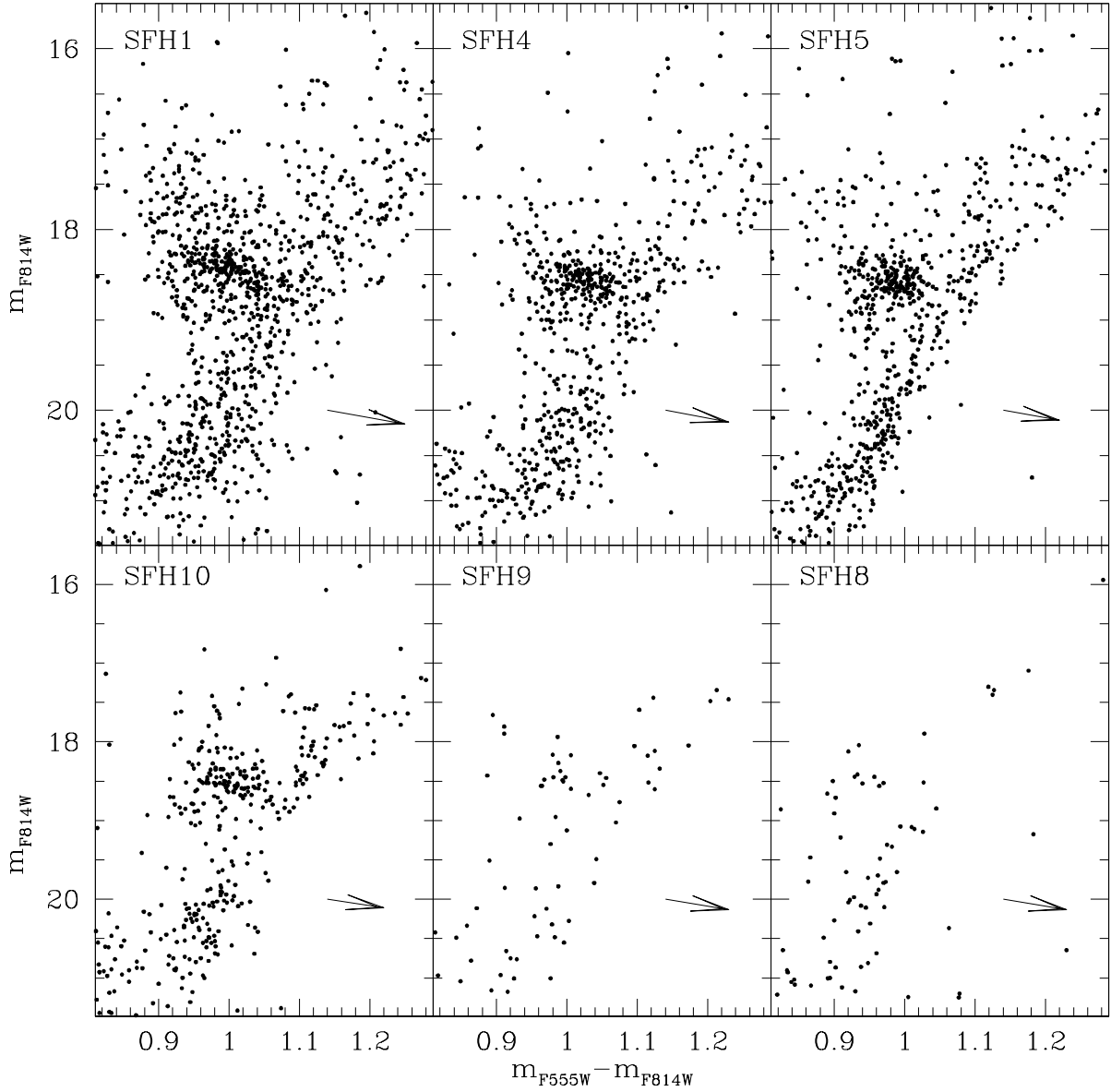}
\caption{\label{f:rcs} CMDs m$_{\rm F814W}$ vs. $m_{\rm F555W}-m_{\rm F814W}$ of the six SMC fields zoomed around the RC. In each plot the arrow on the lower right shows the shift in color and magnitude caused by reddening.}
\end{figure}

Even though RCs are clearly visible in SFH8 and 9, the low number of stars does not allow us to discuss their morphology. Fields SFH1, SFH4, SFH5 and SFH10, on the contrary, show very prominent RCs (Fig.~\ref{f:rcs}), which vary both in shape and color extension. Of these four fields, SFH5 has the bluest RC, with the majority of the stars concentrated between ($0.96 \leq m_{\rm F555w}-m_{\rm F814W}\leq 1.03$), but a fainter and even bluer feature is visible at ($m_{\rm F555w}-m_{\rm F814W}\simeq 0.95, \, m_{\rm F814W}\simeq 18.9$). This feature corresponds to the central helium-burning phase of stars at the low-mass end of the intermediate-mass interval (i.e. M$\sim 2-2.5 M_\odot$), and it has been both predicted and observed in other fields of the SMC and LMC \citep[see for example][]{tosi02}. The RC in SFH10 looks quite similar to that in SFH5 ($0.96 \leq m_{\rm F555w}-m_{\rm F814W}\leq 1.04$, even if the bluer and fainter RC is less visible because of the lower number of stars. The RC is redder in SFH4 ($0.98 \leq m_{\rm F555w}-m_{\rm F814W}\leq 1.08$), suggesting either an increase in the local reddening, or a higher metallicity. In SFH1 the RC has a curved shape and a broad color extension ($0.94 \leq m_{\rm F555w}-m_{\rm F814W}\leq 1.06$), suggesting a spread in age and metallicity, as well as possible differential reddening.

\begin{figure}
\plotone{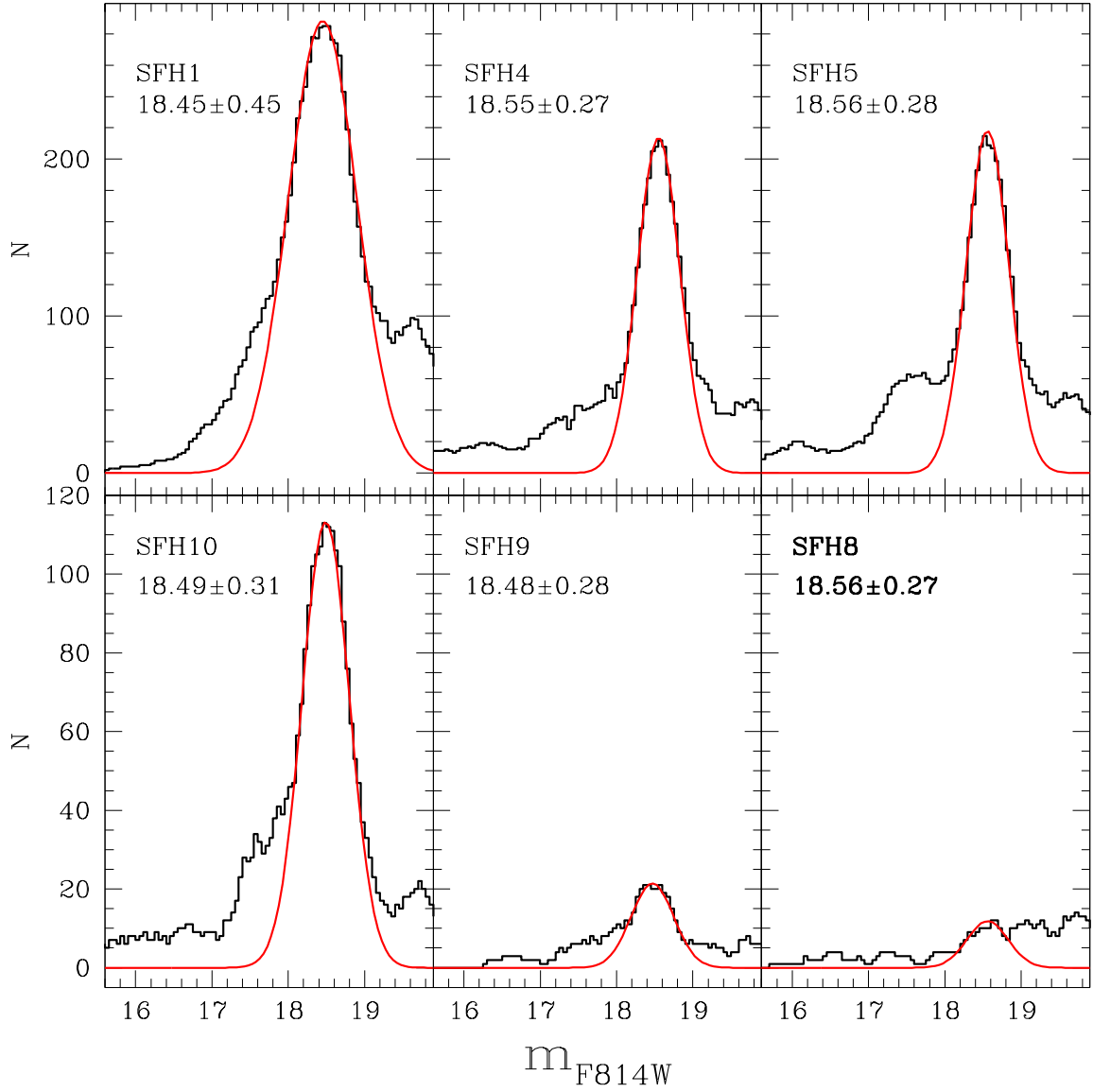}
\caption{\label{f:rc_lf} F814W LFs for the RCs in the six SMC fields studied in this paper. The Gaussian fit (red line) of the RC has been used to derive the mean magnitude of the RC labeled in each panel.}
\end{figure}

In addition to differential reddening and multiple ages, a spread in distance can also cause a broadening of the RC. Different authors have derived rather different distance moduli for the SMC star clusters, even within the same survey (e.g. Mathewson et al. 1988; Hatzidimitriou et al. 1993; Crowl, et al. 2001; Lah, et al. 2005; Glatt et al. 2008b). As a result they concluded that the depth of the SMC can be up to 20 kpc. 
Figure~\ref{f:rc_lf} shows the luminosity function (LF) in the F814W filter for the six RCs. The LFs have been obtained by considering only stars in the color range ($0.8 \leq m_{\rm F555w}-m_{\rm F814W}\leq 1.2$) and brighter than $m_{\rm F814W}< 20$. We used a Gaussian fit to derive mean magnitude. The obtained values are listed in Table~\ref{t:fields}. The dispersion in magnitude (Table~\ref{t:fields}) of the red clump is due to the spread in age, metallicity, reddening, and depth. 

Although not as robust as other standard candles, such as the RGB tip or the period-luminosity relations of pulsating stars, the magnitude of the RC can be used to estimate the distance, once reddening, metallicity and age of the stellar population have been taken into account. To derive the average distance modulus $(m-M)_0$ for each of the observed fields, we used the absolute F814W magnitude ($m_{\rm F814W}=-0.29$) for the RC of a 4 Gyr old stellar population, with metallicity Z=0.001, as predicted by the Padua isochrones \citep{bertelli94}. For each region the value of the derived de-reddened distance modulus is listed in Table~\ref{t:fields}. The mean distance modulus $18.93\pm 0.14$ is consistent with \citet{hilditch05} estimate.

\section{The stellar content of the six SMC fields}
\label{stellar content}

Even if theoretical isochrones do not allow us to put constraints on the intensity of SF episodes, or on their duration, they can be used to infer when major episodes occurred in an observed region. We have therefore superimposed Padua isochrones for different metallicities \citep{bertelli94} onto the six CMDs described before. Isochrones have been transformed to the ACS/WFC photometric system using the transformations by \citet[][and private communication]{origlia}. 
We used the following criteria to fit the observed CMDs:
\begin{itemize}

\item Distance - We used the $(m-M)_0$ derived from the mean F814W magnitude of the RC to determine the average distance for each of the fields discussed in this paper (Table~\ref{t:fields}). 

For ages younger that 10 Gyr the temporal resolution of Padua isochrones is $log(t)=0.1$, then the step becomes $log(t)=0.02$. For stellar populations younger than 10 Gyr therefore the resolution of the model will be the dominant source of uncertainties in determining the age of the stellar population. At older ages the distance modulus becomes the dominant source of error, and will introduce uncertainties of $\sim 1$ Gyr on our estimates. 

\item Metallicity - Two-filter wide-band photometry is not sufficient to infer metal abundances, but it can be still used to give us some hints, with the help of literature information. Recently, \citet{carrera08} derived the metallicity of more than 350 red giants in the SMC using the Ca {\sc ii} triplet. They identified a radial age-metallicity gradient, with the more metal-rich and younger stars being concentrated in the central region of the SMC. They found that stars older than 3 Gyr are more metal-poor than [Fe/H]$=-1$.4, stars between 1 and 3 Gyr have a metallicity [Fe/H]$=-$1.0, and stars younger than $\sim 800$ Myr have even higher metallicity. Spectroscopic analysis of F-type \citep{russell89}, and A-type \citep{venn99} supergiants indicate a mean iron abundance [Fe/H]$=-0.7$. 

Among the available set of Padua isochrones, we therefore selected those that, at the various ages, are closer to the values listed in literature: in particular we used isochrones for metallicity Z=0.004 (corresponding to [Fe/H]$=-$0.7 in the scale of Anders \& Grevesse 1989, [Fe/H]$=-$0.64 in the Grevesse \& Sauval 1998 scale, and [Fe/H]$=-$0.5 in the Asplund 2004 scale) to fit the young stellar population, and Z=0.001 ([Fe/H]$=-$1.3 in Anders \& Grevesse 1989 scale, [Fe/H]$=-$1.25 in the Grevesse \& Sauval 1998 scale, and [Fe/H]$=-$1.1 in the Asplund 2004 scale) for the intermediate and old age stellar populations, except when different metallicities were clearly needed. 

\item Reddening - Different regions of the SMC exhibit different amounts of dust and gas content \citep{schlegel98}: in the SMC bar, for example, dust and gas are more abundant than in the external regions \citep{staveley97,stanimirovic99}. Therefore we expect to find different reddening values in our fields.

To put a lower limit to the reddening values, we used the blue side of the upper MS (UMS), i.e. the MS brighter than $m_{\rm F814W}<22$ and bluer than $m_{\rm F555W}-m_{\rm F814W}<0.3$. This estimate depends on the isochrones chosen to infer the value, and has uncertainties of $\pm 0.01$. In some regions (see discussion below) we found evidence of differential reddening up to the $\sim 50$\%

\end{itemize}

In a forthcoming paper (Cignoni et al., in preparation) we will use the synthetic CMD technique to infer the SFH of each field. This method will allow us also to quantify the impact of the uncertainties listed above on our estimates.

\subsection{The Bar}
\label{bar}

\begin{figure}
\plotone{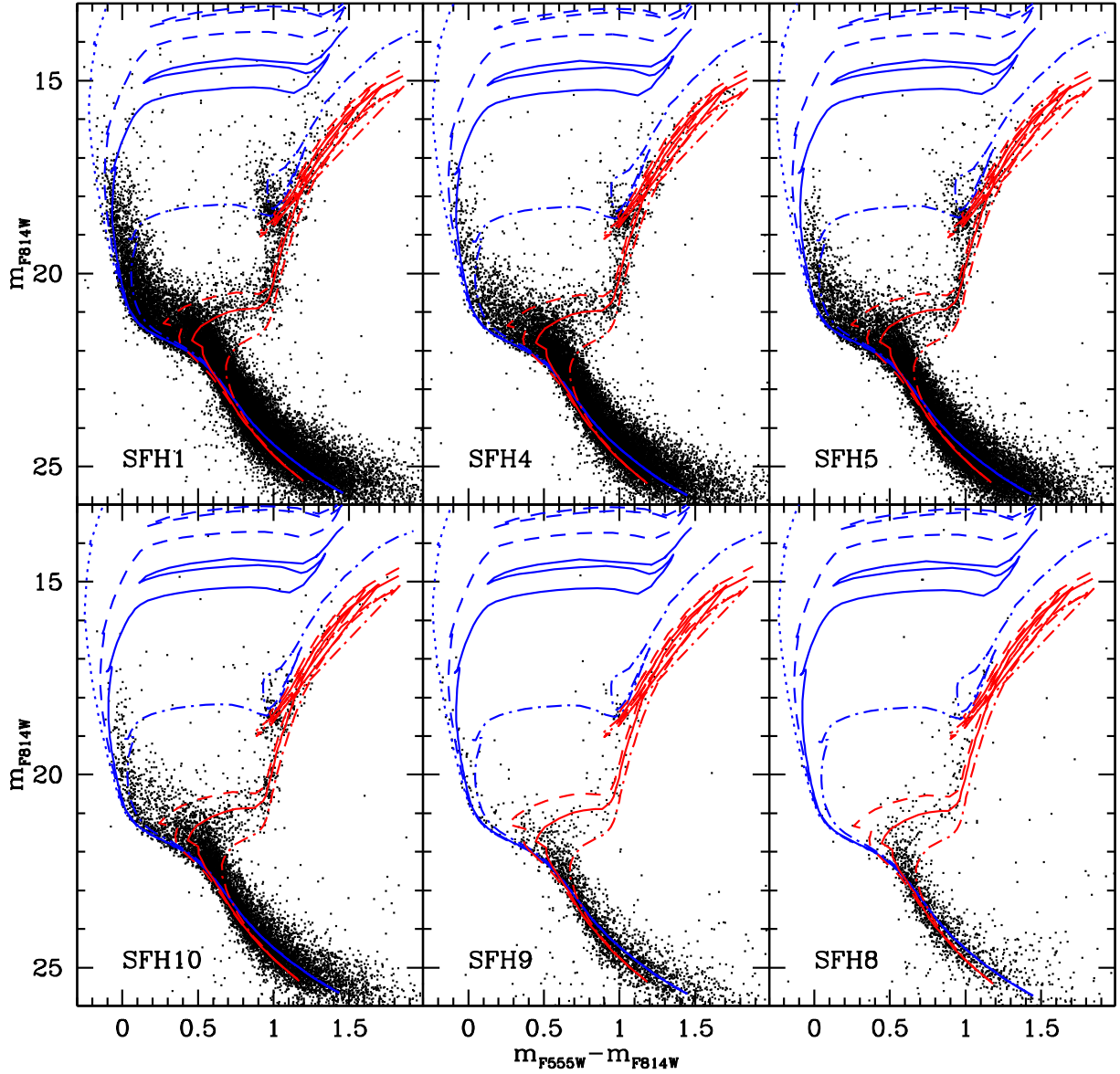}
\caption{\label{f:ages} CMDs m$_{\rm F814W}$ vs. $m_{\rm F555W}-m_{\rm F814W}$ of the six SMC fields with superimposed Padua isochrones of metallicity Z=0.004 (in blue) Z=0.001 (in red). Red dotted-dashed lines are isochrones for 12 Gyr, continuous red isochrones are for 5 Gyr, and red dashed isochrones are for 3 Gyr old stellar populations. Blue dotted-dashed lines are 500 Myr isochrones, continuous blue isochrones are for 100 Myr, dashed blue isochrones are for 50 Myr and dotted blue isochrones are for 10 Myr old stellar populations.} \end{figure}

Figure~\ref{f:ages} shows the $m_{\rm F814W}$ vs. $m_{\rm F555W}-m_{\rm F814W}$ CMDs of the six SMC regions with a selection of Padua isochrones superimposed. In the bar we examined three different regions (Figure~\ref{f:targets}): SFH1, SFH4 and SFH5.

\subsubsection{SFH1}

In our dataset SFH1 is the region closest ($\sim 24\arcmin$) to the optical center of the SMC, and characterized by the highest stellar density \citep[$\sim 0.73\, {\rm stars\, arcsec}^{-2}$, corresponding to $\sim 8.5\, {\rm stars\, pc}^{-2}$ at a distance of 60.6 kpc][]{hilditch05}. 

In order to reproduce the bluest colors of the UMS in SFH1 we adopted a reddening of $E(B-V)=0.08$, however the broad upper MS suggests that locally the reddening can vary up to $\sim 50$\%. This variation is not sufficient to explain the thickness of the RGB suggesting also a spread in metallicity at intermediate ages. 

\begin{figure}
\plotone{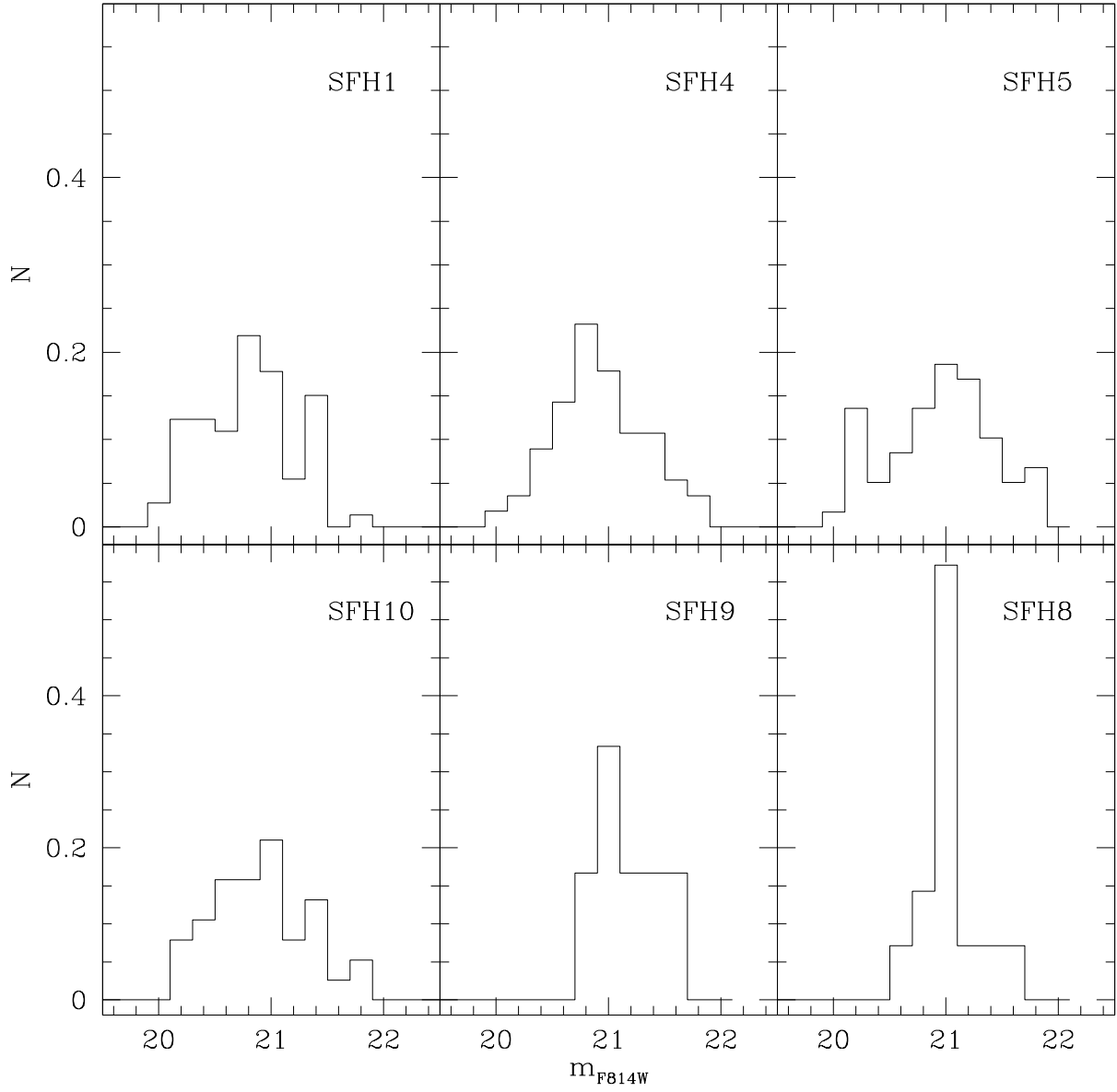}
\caption{\label{f:lf_sgb} SGB-LF of the observed regions. The histograms have been normalized to the total number of stars plotted.}
\end{figure}

The broad SGB ($20.0\la m_{\rm F814W} \la 21.5$) suggests that SFH1 formed the majority of the stars between $\sim 1.5$ and $\sim 12$ Gyr ago. Such a prolonged SF activity is also supported by the complex shape of the RC. Figure~\ref{f:lf_sgb} shows the luminosity function (LF) of the SGB obtained from stars in the magnitude range $20<m_{\rm F814W}<21.8$ and color range $0.8<m_{\rm F555W} - m_{\rm F814W}<0.9$. In SFH1 (upper left panel) the SGB-LF is peaked at $m_{\rm F814W}=20.8$, that, assuming a metallicity Z=0.001, corresponds to $\sim 4$ Gyr. As mentioned in \S~\ref{cmds} we find only a few stars with magnitudes and colors consistent with an HB, suggesting that in the first few billion years SF in this region was low. This is also supported by the small number in the SGB-LF of faint ($m_{F814W}<=21.4$) SGB stars ($\sim 16$\%) older than $\sim 8$ Gyr. 

We found that Z=0.001 isochrones 2.5 Gyr old predict colors that are too blue for both the MSTO and the RGB, suggesting that the interstellar medium was already relatively enriched at that time. 

We used Z=0.004 isochrones to fit stars younger than 1 Gyr. The well-populated UMS suggests that in SFH1 stars were probably formed continuously from $\sim 800$ Myr ago to $\sim 100$ Myr ago. 

The concentration of He burning stars above the RC ($16.0\la m_{\rm F814W} \la 18.0 \, m_{\rm F555W}-m_{\rm F814W}\simeq 0.9$) corresponds to the lower end of the blue edge of the blue loop. These stars formed between 400 and 500 Myr ago. 
In order to quantify the relative intensity of the SF occurring $\sim 500$ Myr ago among the six field, we considered the ratio between the He burning stars above the RC and the stars in the RC (BL/RC). For this test we considered as blue loop stars those in the magnitude $16.0< m_{\rm F814W} < 18.0$ and color ranges $0.7< m_{\rm F555W}-m_{\rm F814W} < 1.05$, while we counted as RC stars those between $18.0< m_{\rm F814W} < 18.9$ and $0.9< m_{\rm F555W}-m_{\rm F814W} < 1.1$. The average BL/RC=0.2, while in SFH1 it rises up to BL/RC=0.4; thus $\sim 500$ Myr ago this was the most actively star forming region in our survey. 

The evolved bright stars with intermediate colors ($m_{\rm F814W} \la 16.7, \, 0.2 \la m_{\rm F555W}-m_{\rm F814W}\la 0.9$) indicate that SF has been active in the last 400 Myr, however since the bright blue loop is not as well defined as above the RC, it is possible that the star formation rate (SFR) has decreased over the last few hundred Myr. The very blue ($m_{\rm F555W}-m_{\rm F814W}< 0.0$) and bright ($m_{\rm F814W}< 15.0$) stars indicate that SF was still ongoing 20-30 Myr ago, in agreement with the circumstance that we are watching a region close to a very active part of the SMC. 

\subsubsection{SFH5}

SFH5 is the second densest region of our survey ($\sim 0.5\, {\rm stars\, arcsec}^{-2}$ corresponding to $\sim 5.7\, {\rm stars\, pc}^{-2}$). The best fit to the blue edge of the UMS is obtained by adopting a reddening E(B-V)=0.06. Of our six regions, this is the one with the best-defined (tight and well populated) evolutionary sequences. 
The RGB is tighter than that in SFH1, suggesting a smaller range of metallicities, or a less variable reddening. This is also supported by the smaller color spread of the RC (Figure~\ref{f:rcs} -- upper right panel), and by the group of stars fainter and bluer than the clump, corresponding to the lowest mass stars (M$\sim 2 M_\odot$) that could ignite helium nuclear reactions in a non degenerate core. They represent the faint vertex of the blue loops, and are better recognizable than in SFH1 and in the other fields, thanks to the tighter evolutionary sequences. 

The SGB-LF shows that, as in SFH1, SFH5 formed only a few stars ($\sim 22$\%) between 12 and 8 Gyr ago. The majority of the stars are older than $\sim 2$ Gyr, and, as in SFH1, the SGB-LF is peaked at $m_{\rm F814W}=21$ which suggests an enhancement in SF occurred $\sim 5$ Gyr ago. As in SFH1, we find evidence of chemical enrichment with time: the Z=0.001 isochrones nicely fit the oldest stellar populations, but predict colors bluer than observed for stars younger than $\sim 3$ Gyr.

Stars in the blue loop above the RC ($17\la m_{\rm F555W} \la 18.2$) suggest that SF increased $\sim 500$ Myr ago. The youngest stars in this region are likely $\sim 200$ Myr old, implying low levels of SF activity in the most recent epochs. 

\subsubsection{SFH4}

The third bar field of our sample for stellar density ($\sim 0.4\, {\rm stars\, arcsec}^{-2}$; $\sim 5\, {\rm stars\, pc}^{-2}$) is SFH4. Figure~\ref{f:ages} (upper-central panel) shows its $m_{\rm F814W}$ vs. $m_{\rm F555W}-m_{\rm F814W}$ CMD with Padua isochrones superimposed. We assumed $E(B-V)=0.07$ to fit the blue edge of the UMS. The oldest MSTO ($m_{\rm F814W}\simeq 22.1, \, m_{\rm F555W}-m_{\rm F814W} \simeq 0.6$) and faintest SGB indicate that SF was already occurring $\sim 12$ Gyr ago. The best fit is achieved assuming a metallicity Z=0.001. The broad SGB ($20.2\la m_{\rm F814W} \la 22.1$) indicates that SF went on until $\sim 2$ Gyr ago. The SGB-LF (Figure~\ref{f:lf_sgb} upper-central panel) is peaked at $m_{\rm F814W}=20.8$, suggesting that a possible increase in SF occurred $\sim 4$ Gyr ago. As in SFH1, the SGB-LF shows that only few stars ($\la 20$\%) are fainter than $m_{\rm F814W}=21.4$ and thus older than $\sim 8$ Gyr. Z=0.001 isochrones younger than 3 Gyr predict too blue colors for the MS and RGB, suggesting a higher metallicity. 

The most recent episode of SF likely occurred between $\sim 50$ and 200 Myr ago, while the blue loop stars above the RC suggest a moderate activity about $\sim 500$ Myr ago (BL/RC=0.13).

The comparison of the three innermost regions shows that the SMC formed stars in the first 10-12 Gyr with a possible major episode occurring between 6 and 4 Gyr ago. We found a progressive chemical enrichment with stars older than 2-3 Gyr being more metal poor (Z=0.001), than stars younger than $\sim 1$ Gyr (Z=0.004). 

\begin{figure}
\plotone{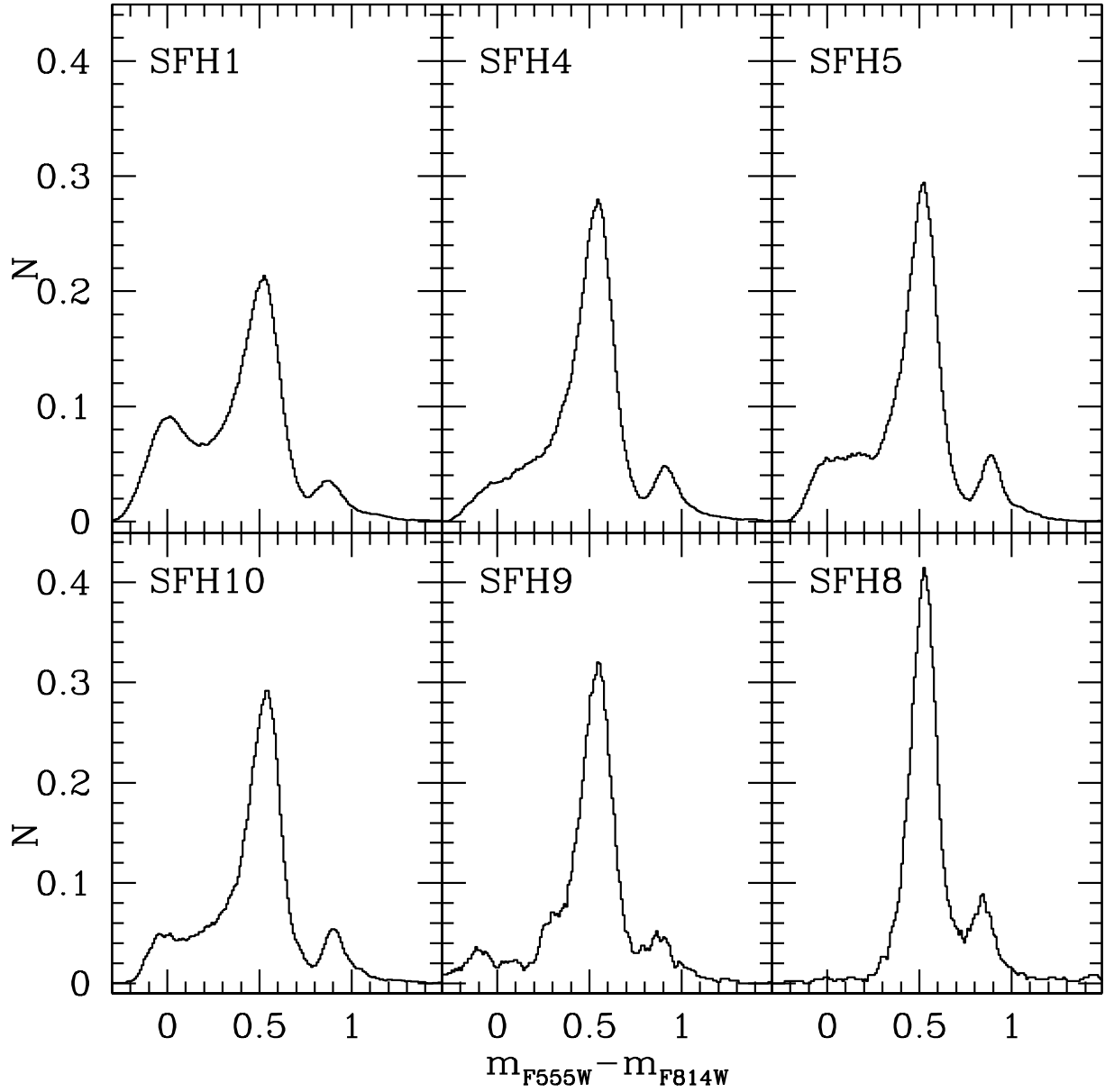}
\caption{\label{f:cf} CFs for the observed regions, obtained by considering only the stars brighter than the oldest MSTO.}
\end{figure}

Isochrone fitting does not allow us to infer any information about the SFR, but we can combine it with the color function (CF) to highlight relative differences in the intensity of star-forming episodes. A CF plots the number of stars of different colors. Figure~\ref{f:cf} shows the CFs obtained for the six SMC regions discussed in this paper. In each plot we considered only stars above the oldest MSTO ($m_{\rm F814W}<23.0$). This selection guarantees that bias effects due to incompleteness are negligible. In each plot the CF has been normalized to the total number of stars brighter than $m_{\rm F814W}<23.0$, and corrected for the reddening value derived from the isochrone fitting to the UMS described above.

In all the CFs of  Figure~\ref{f:cf} we notice that the majority of the stars have $0.28\leq m_{\rm F555W}-m_{\rm F814W}\leq 0.75$. The peaks of all fields coincide with each other at $m_{\rm F555W}-m_{\rm F814W}=0.53$, thus confirming our estimate of the relative differences in the local reddening values. This peak is mostly populated by old and intermediate age stars in MS and SGB evolutionary phases. In particular, it is the locus of the low-mass (M$\leq 1.2 M_\odot$) MSTO stars.

The redder ($m_{\rm F555W}-m_{\rm F814W}\geq 0.75$) secondary maximum in the CFs corresponds to evolved stars: the end of the SGB, RGB, RC, AGB, and red supergiants. In our cases the main contribution comes from SGB, RC and RGB stars, as such this peak is representative of a stellar population older than 1-2 Gyr. This color of peak is sensitive to age and metallicity.

The bluer side ($m_{\rm F555W}-m_{\rm F814W}\leq 0.28$) of the CFs contains intermediate and massive stars in MS and at the blue edge of the blue loop evolutionary phases, and is representative of a stellar population younger than $\sim 800$ Myr. The relative strength of the secondary blue peak with respect to the red one gives an indication of the fraction of young to old stars. Hence, we can use the relative intensity of the peaks to infer when a region was more active with respect to another. For instance, we found that in SFH1 33\% of the stars brighter than $m_{\rm F814W}<23$ are bluer 
(younger than 800 Myr), while in SFH4 only 19\% of the stars have such blue colors. In SFH5 the fraction of blue stars is 21\%.

We also note that the morphologies of the blue side of these three CFs are quite different: SFH1 presents a clear peak at $m_{\rm F555W}-m_{\rm F814W}\simeq 0.03$, while in SFH4 we observe a smooth decrease in stellar counts. The CF in SFH5 shows a plateau between $0\leq m_{\rm F555W}-m_{\rm F814W}\leq 0.22$ and also has a shorter color extension toward the blue compared to the other two regions. This comparison again suggests that in the last billion years SFH1 was the most active region, and that a sizeable fraction of its stars were formed $\sim 400-500$ Myr ago. This region continued to form stars until $\sim 20-30$ Myr ago, but at a lower rate. SFH4 formed stars probably at a steadier pace than SFH1. Finally, the younger stellar population of SFH5 formed mainly between 200 and 400 Myr ago.

\subsection{The Wing}
\label{wing}

The tidal interaction between the Magellanic Clouds is considered to be at the origin of the wing. From the analysis of a small portion of the SMC wing, \citet{mccumber05} concluded that stars in the wing formed during a prolonged episode of SF between 3.5 and 12 Gyr ago. Their MS is well populated up to V=17.8, indicating very recent star forming activity. From isochrone fitting they concluded that the youngest stars can not be older than 100 Myr, and other episodes of SF may have occurred $\sim 220$ Myr, $\sim 1.3$ Gyr, $\sim 1.8$ and $\sim 2.2$ Gyr ago. They also noted that the 100 Myr isochrones was not as densely populated as the rest of the young MS, and concluded that the SFR in the wing is currently at lower levels compared to the recent past.  

We analyzed the stellar content of two regions of the wing, one closer to the bar (SFH10) and one towards the bridge (SFH9) (see Figure~\ref{f:targets}).

\subsubsection{SFH10}
SHF10 is located in the central part of the wing, at $\sim 1\degr 36\arcmin$ from the H{\sc ii} complex N83-N84-N85, which is considered an example of SF triggered by cloud collisions, and at less than $8\arcmin$ from the intermediate age star cluster NGC~419 with its complex, seemingly multiple MSTOs \citep[1.2-1.6 Gyr ][]{glatt08b}. 
In SFH10 the stellar density is approximately one third of that in SFH1 ($\simeq 0.3\, {\rm stars\, arcsec}^{-2}$ or $\sim 2.7\, {\rm stars\, pc}^{-2}$).

To superimpose the Padua isochrones on the CMD of SFH10, we need a reddening value $E(B-V)=0.06$. As for the bar, we used Z=0.001 isochrones to fit the older stellar population and Z=0.004 for the younger stars (Figue~\ref{f:ages} - bottom-right panel). Comparison with the isochrones shows that also in the wing the SF activity was already in place $\sim 12$ Gyr ago. The width of the SGB and its smoothness suggest that SFH10 formed stars at a steady pace between $\sim 5$ and $\sim 3$ Gyr ago. The peak of SGB-LF is at $m_{\rm F814W}=21$, which corresponds to $\sim 5$ Gyr, $\sim 20$\% of the stars are older that 8 Gyr ($m_{\rm F814W}>21.4$. We also noted that for ages younger than $\leq 3$ Gyr, the Z=0.001 isochrones predict bluer than observed colors for the MSTOs and RGBs, indicating a metallicity higher than in the models. The majority of the stars in the UMS were formed likely between 300 and 500 Myr ago, and few stars can be fitted with a $\sim 100$ Myr old isochrone, in agreement with the \citet{mccumber05} conclusions that the wing was forming more stars in a recent past than now.

\subsubsection{SFH9}
SFH9 is in a more external region of the wing and is also the pointing in our survey most distant from the optical center of the SMC ($>7\degr$). The stellar density drops down to $\simeq 0.07\, {\rm stars\, arcsec}^{-2}$ ($\sim 0.8\, {\rm stars\, pc}^{-2}$) in SFH9. 

The majority of the stars in the SFH9 CMD are associated with the old and intermediate age stellar population, with MS, SGB, RGB and RC reasonably well defined. As done for the other fields, we have used the colors of the UMS to constrain the reddening value, and obtained the best agreement with the observed data for $E(B-V)=0.07$. 
The older stellar population is composed of stars formed mainly between 4 and 10 Gyr ago.

\begin{figure}
\plotone{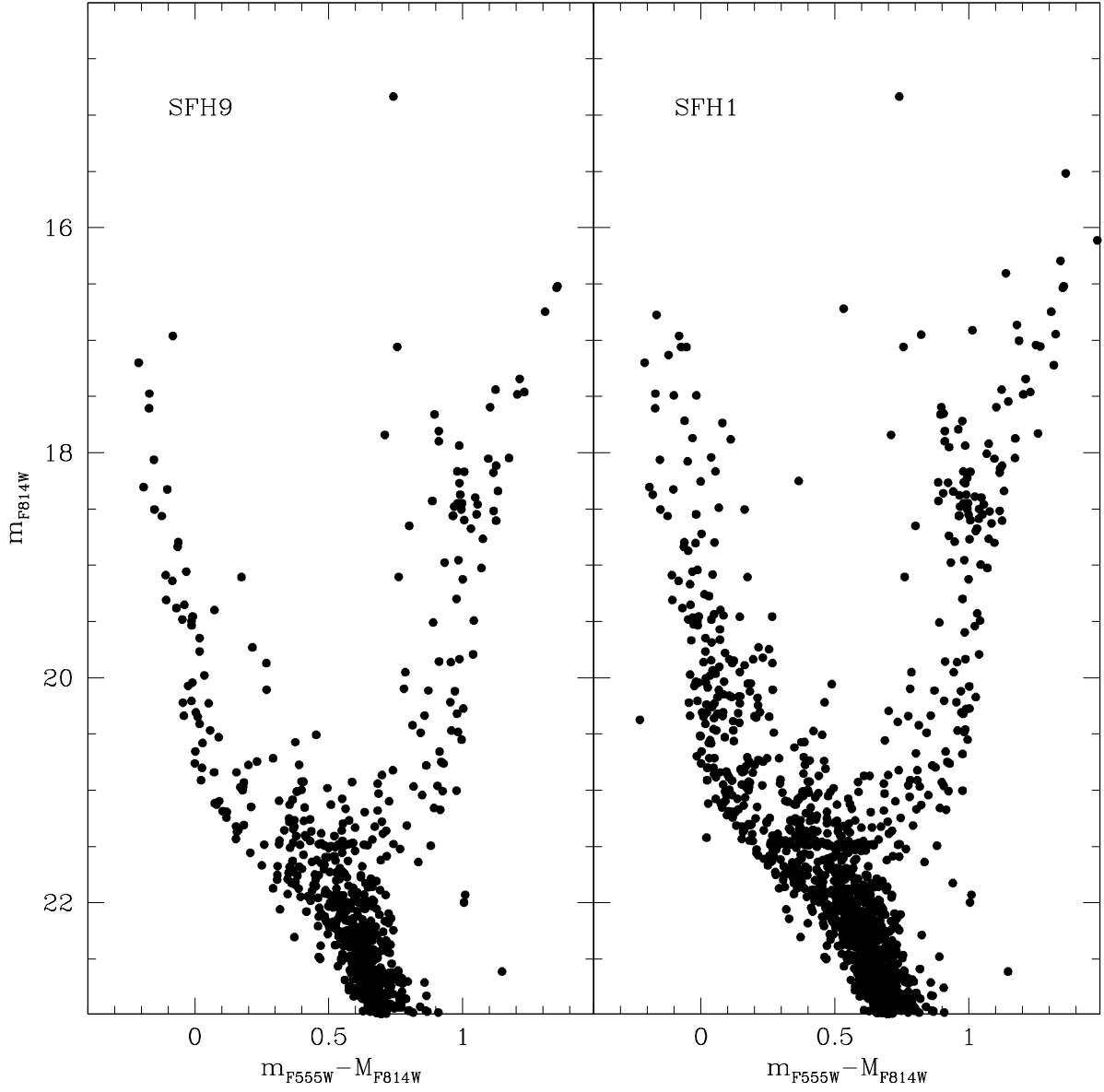}
\caption{\label{f:gdc} {\it Left Panel} SFH9 CMD for stars brighter than $m_{\rm F814W}< 23$. 700 stars are plotted. {\it Right Panel}: CMD of 700 stars with magnitude $m_{\rm F814W}<23$ randomly extracted from SFH1.} 
\end{figure}

The UMS is very blue and, at all magnitudes, very tight, suggesting that nearly all these stars are younger than $\sim 100$ Myr. Preliminary tests indicate that it is unlikely that the narrowness of the UMS is due only to a poor statistics. In SFH9 there are $\sim 700$ stars brighter than $m_{\rm F814W}<23.0$. We have thus randomly extracted $\sim 700$ stars brighter than $m_{\rm F814W}<23.0$ from SFH1 for 50 times. Figure~\ref{f:gdc} shows the comparison between the CMD of SFH9 (left panel) and one of the sub-samples randomly extracted from SFH1 (right panel). We found that in SFH9 the stars fainter than $m_{\rm F814W}>19.5$ in the color range $-0.1<m_{\rm F555W}-m_{\rm F814W}<0.3$ are on average only 33$\pm 3$\% of those found in SFH1. We also compared the spread in colors between the ridgeline of the UMS of SFH9 and all the stars bluer than $m_{\rm F555W}-m_{\rm F814}<0.3$ both in SFH9 and in the sub-sample of stars extracted from SFH1. This test shows that the UMS of SFH9 is tighter than SFH1 with a 4$\sigma$ significance.

\subsection{The SMC outskirts}
\label{halo}

SFH8 was selected to sample the stellar content in the outskirts of the SMC, which may be representative of the SMC stellar population formed before the episodes of star formation that shaped its present day irregular morphology. SFH8 is $> 3\degr$ North from the SMC optical center, and has a stellar density of only $\simeq 0.04\, {\rm stars\, arcsec}^{-2} (\sim 0.5\, {\rm stars\, pc}^{-2}$). 

\begin{figure}
\plotone{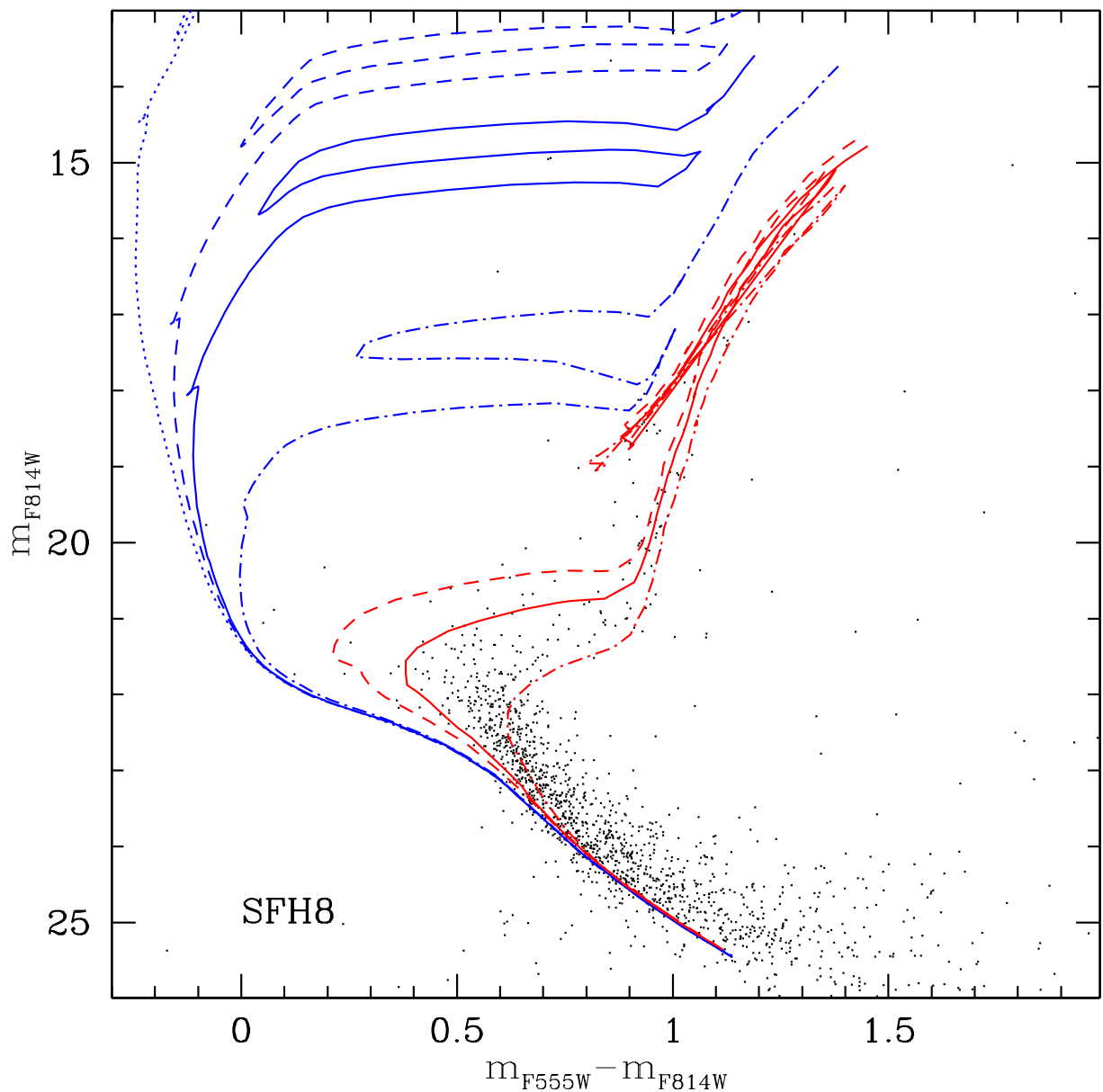}
\caption{\label{f:sf8} CMD m$_{\rm F814W}$ vs. $m_{\rm F555W}-m_{\rm F814W}$ of SFH8 with superimposed Padua isochrones of metallicity Z=0.0004. Red dotted-dashed lines are isochrones for 12 Gyr, continuous red isochrones are for 5 Gyr, and red dashed isochrones are for 3 Gyr old stellar populations. Blue dotted-dashed lines are 500 Myr isochrones, continuous blue isochrones are for 100 Myr, dashed blue isochrones are for 50 Myr and dotted blue isochrones are for 10 Myr old stellar populations.} \end{figure}

The lower-right panel in Figure~\ref{f:ages} shows the $m_{\rm F814W}$ vs. $m_{\rm F555W}-m_{\rm F814W}$ CMD with Z=0.004 (blue) and Z=0.001 (red) Padua isochrones. The RGB is bluer and less curved than the Z=0.001 isochrones and, as shown in Figure~\ref{f:sf8}, at all the ages we obtained a better fit using very metal poor isochrones (Z=0.0004, corresponding to [Fe/H]$=-1.82$ using the solar abundances by Anders \& Grevesse 1989, [Fe/H]$=-1.76$ in the scale by Grevesse \& Sauval 1998 and to [Fe/H]$=-1.62$ in the Asplund 2004 scale). Also the CF confirms the lower metallicity of this region, since it shows the bluest red peak in our sample. The low metallicity of SFH8 is consistent with the Ca {\sc ii} measurements by \citet{carrera08}, who found stars as metal poor as [Fe/H]$=-1.64$ in the outer regions. \citet{carrera08} noted a radial metallicity gradient that can explain the differences between the stellar populations found in SFH8 and the region observed by \citet{dolphin01}, who, observing a region $\sim 2\degr 25\arcmin$ from the center of the SMC (and $1\degr 52\arcmin$ from SFH8), found evidence of progressive chemical enrichment with age from Z=0.0008 to Z=0.004

With the exception of the 14 objects bluer than $m_{\rm F555W}-m_{\rm F814W}\le 0.4$, assuming a metallicity Z=0.0004, all the stars in the CMD are likely older than $\sim 5$ Gyr, with the majority formed between 5 and 10 Gyr. 

The 14 blue objects that mimic an extension of the MS to brighter magnitudes can be fit with a $\sim 2$ Gyr old isochrone, and might be, as suggested by \citet{dolphin01} for their halo field, the evidence of a very small episode of star formation. Alternatively, they could be blue straggler stars \citep[BSS][]{sandage53}. Since, according to the Padua isochrones, the bluest and brightest star ($m_{\rm F555w}-m_{\rm F814W}\simeq -0.08;\, m_{\rm F814W}\simeq 19.8$) in the UMS has a mass of $M\sim 1.5\, M_\odot$, while the MSTO of a 5 Gyr old stellar population has a turn-off mass of 1.06 $M_\odot$, the BSS hypothesis is not inconceivable. We will test both hypotheses with the help of synthetic CMDs in a forthcoming paper (Cignoni et al. in preparation).

Finally we note that the stellar populations in SFH8 clearly differ from those commonly observed in the halo of the MW, in the sense that the MW halo is dominated by stars older than 10 Gyr, whereas only few such stars are in the SMC halo. 

\subsection{Comparison with some SMC star clusters}

\begin{figure}
\plotone{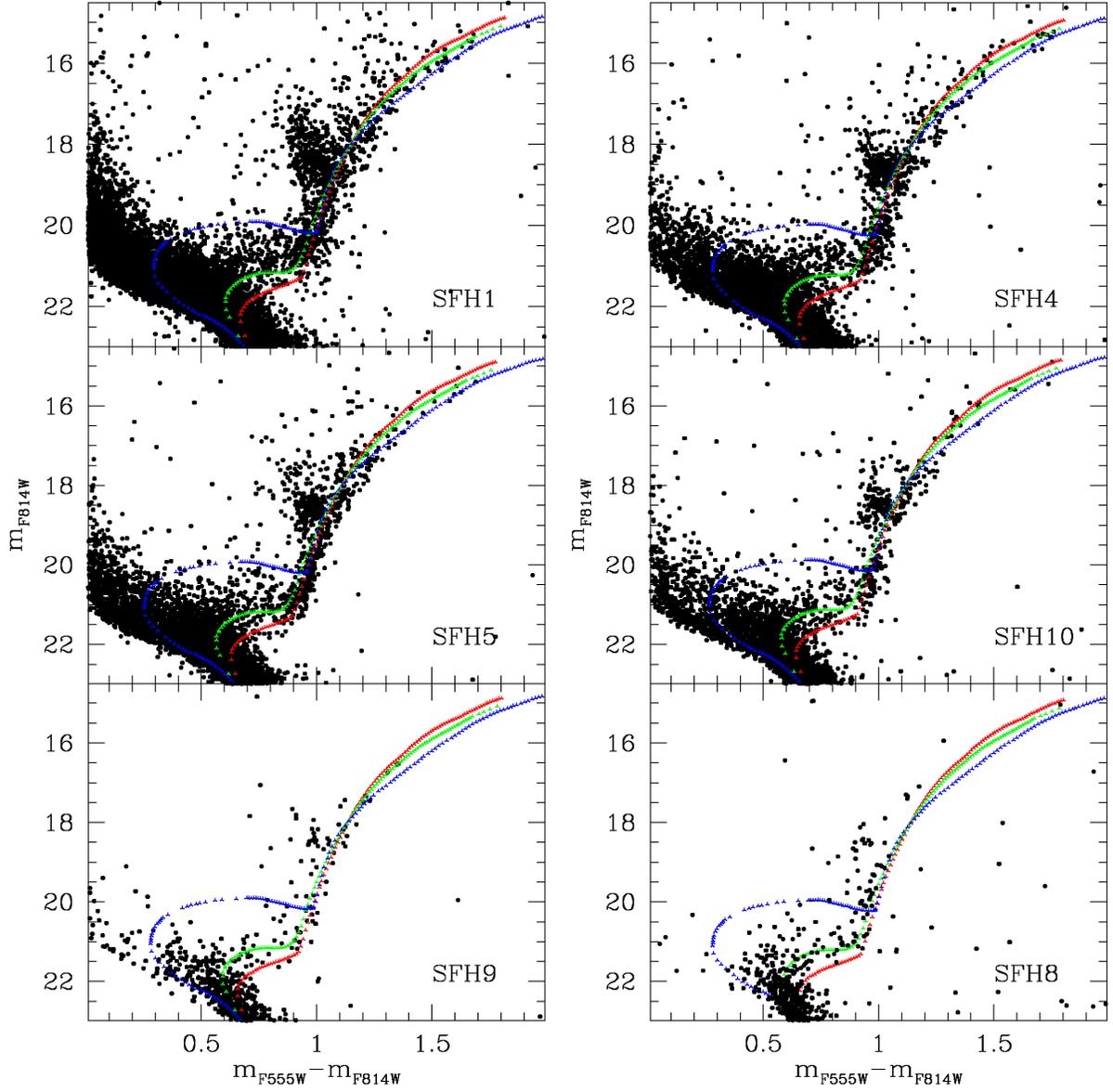} 
\caption{\label{f:clusters} CMDs m$_{\rm F814W}$ vs. $m_{\rm F555W}-m_{\rm F814W}$ of the six SMC fields zoomed on the loci of the evolved old stars. Ridgelines for NGC~121, Kron~3 and NGC~419 are shown in red, green and blue respectively.}
\end{figure}

\citet{glatt08b} pointed out that several theoretical models fail to reproduce the upper RGB slope, a feature often used in broadband photometry to infer the metallicity of intermediate and old stellar populations, and the SGB morphology. To further characterize the properties of the old stellar component in the SMC field, in Figure~\ref{f:clusters} we therefore superimposed to the CMDs of the SMC fields discussed in this paper, the ridgeline of three of the star clusters published by \citet{glatt08a,glatt08b}. We chose NGC~121 as a prototype of a very old ($\sim 11.5$ Gyr) very metal poor (Z=0.0006) stellar population, Kron~3 as representative of an old ($\sim 6.9$ Gyr) metal poor (Z=0.001) one, and NGC~419 to represent a relatively young ($\sim 1.2$ Gyr) chemically enriched (Z=0.004) stellar population.

In all the observed fields the NGC~121 ridgeline (red line in Figure~\ref{f:clusters}) nicely reproduces the morphology of the lower SGB confirming the presence of an old (age $>10$ Gyr) and very metal poor stellar population in the SMC. It also overlaps the narrow sequence of stars bluer than the main RGB. Even if it is likely that this feature is dominated by low mass metal poor AGB stars, some of these stars can be very metal poor (Z$\la 0.0006$) bright RGB stars. 

The ridgeline of NGC~419 (blue line) fits very well the redder component of the upper RGB in SFH1, 4, 5, and 10, confirming that both in the bar and the wing the metallicity was close to the present day SMC metallicity already few Gyr ago. In these four CMDs only a few stars have colors and magnitude consistent with $\sim 1$ Gyr old stellar population, further suggesting that star formation occurred at a lower rate than in the previous epochs.

Finally it is worth noting that stellar fields and clusters in the SMC seam to have had a similar history of formation, with the majority of stars and clusters      
\citep[e.g.][]{olszewski96,dacosta,glatt08a,glatt08b} formed between $\sim 4$ and $\sim 7$ Gyr ago. This is opposite to what we observe in the LMC, where clusters formed in two main episodes \citep[occurred $\sim4 $ and $\sim 12$ Gyr ago][]{vandenberg91,bertelli94}, while the SFH in the field was fairly continuous \citep{holtzman99}  

\section{Discussion and conclusions}
\label{conclusions}

The SMC is the closest dwarf galaxy with ongoing star formation, but until recently it has received less attention than the LMC. As mentioned in the Introduction, only a few groups have studied the SFH of its field stars, \citet{dolphin01}, \citet{harris04}, \citet{mccumber05}, \citet{chiosi07}, \citet{noel07}, and \citet{cignoni09}. Our survey is designed to help fill this gap. The six fields we observed with the HST/ACS sample SMC regions 
characterized by very different star and gas densities, and, possibly, by different evolutionary histories. In all the six regions our photometry goes $\ga 3.5$ magnitudes below the oldest MSTO, allowing us to investigate in detail the properties of stars formed over the entire Hubble time. Our survey is currently the deepest and has the highest spatial resolution ever performed for the SMC fields, and allows us to investigate even its densest regions, such as the bar.

Our data show that the SMC was already forming stars $\sim 12$ Gyr ago, yet we did not find a well-defined HB, i.e. the signature of an old and metal-poor stellar population, in any of our fields. On the other hand, a comparison with the ridgeline of SMC star clusters shows that old ($\sim 12$ Gyr) and metal-poor (Z$\la 0.0006$) stars are likely present in every field. These circumstances suggest that in the first few billion years the SF activity in the SMC was low. This is also confirmed by the overall small number ($\sim 2\times 10^{-5}$ arcsec$^{-2}$) of RR-Lyrae found in the SMC bar \citep{soszynski02}. If fields and star clusters in the SMC have similar SFHs, this can also explain why NGC~121, the only globular cluster in the SMC, is a couple of billion years younger than the globulars in the MW and LMC \citep{glatt08a}.

The hypothesis of a slow start of the SF activity is at odds with the conclusion by \citet{harris04} that the SMC formed approximately 50\% of all the stars more than 8.4 Gyr ago, and only few stars between 3 and 8.4 Gyr ago. On the other hand, our result is consistent with that of \citet{noel07}, who did not find the HB in any of their 12 fields, and concluded that the SMC formed very few stars at the beginning, experiencing an intense and prolonged SF activity between 1 and 6 Gyr ago. \citet{chiosi07} also found that the SF activity in the field population presents periods of enhancement 0.3-0.4, 3-4 and 6 Gyr ago, and a likely quiescence at earlier epochs, as if early on, the tidal interactions between the Clouds were not able to trigger any significant SF. A negligible SF activity at the earliest epochs was also found by \citet{cignoni09} in the NGC~602 region in the wing. Future surveys devoted to identify all the variable stars in the SMC \citep[i.e. with the ESO survey telescopes VISTA, in the IR, and VST, in the optical bands; see][respectively]{cioni08,ripepi06}
will provide a complete census of the RR-Lyrae in the whole SMC, thus allowing to fully trace the spatial distribution of oldest populations. This will also allow us to establish whether any of the stars in our fields is an RR-Lyrae, and therefore belongs to the HB.

From our CMDs we infer that the SMC formed stars over a long interval of time until $\sim 2-3$ Gyr ago. With the exception of the external field SFH8, intermediate-age stars show rather uniform characteristics, and a possible increase in the star forming activity approximately between 4 and 6 Gyr ago, in agreement with \citet{noel07} and \citet{chiosi07}. Also SFH8 shows evidence of prolonged star formation in the past, but the lower metallicity of this region suggests that SF might have occurred at a lower rate. 

According to cold dark matter simulations hierarchical structures should form at all scales \citep{diemand07}, therefore if the MCs formed outside the MW halo \citep{donghia08} and now are at their first passage near the Sun \citep{kallyvayalil06, besla07}, they should be still surrounded by their own stellar halo. An extended dynamically hot stellar-like halo has been indeed found around the LMC \citet{majewski00,majewski09}, and stars that belong to the SMC have been found up to $\sim 6.5$ kpc away from this galaxy. Is this extended component part of a MC halo? A systematic spectro-photometric analysis of the surroundings of the SMC is needed not only to verify the existence of an extended halo, and characterize its properties, but also to provide new insights in the formation and evolution of the entire LG. 

We find that stars younger than 800 Myr are concentrated in the bar and the wing of the SMC. In the inner fields (SFH1, SFH4, SFH5 and SFH10), we see an enhancement in the SF activity starting from approximately 500 Myr ago, again in agreement with \citet{chiosi07}. Stars younger than $\sim 100$ Myr have a very inhomogeneous distribution, indicating that recent SF has locally developed.


Within the uncertainties of our two-band photometry, we found evidence of radial chemical enrichment, with the most external regions (the outskirts and the external wing) characterized by lower metallicity than the central zones. This result is in agreement with that by \citet{carrera08} and suggests that the SF activity was significantly lower in the periphery of the SMC than in the center. 
A weak gradient of decreasing metallicities at larger distance has been also found in old and intermediate--age clusters \citep{glatt08b}.

Stellar fields and clusters seems to have had a similar history of formation in the SMC, opposite to what we observe in the LMC, where clusters formed in two main episodes \citep[occurred $\sim4 $ and $\sim 12$ Gyr ago][]{vandenberg91,bertelli94}, while the SFH in the field was fairly continuous \citep{holtzman99}  

The qualitative analysis of the CMDs presented here has already led to some interesting insights into the SFH of the SMC . In a future paper we will perform a more quantitative analysis of the SFH in the six regions, fully exploiting all the photometric information, by applying the synthetic CMD method, which takes into account photometric errors, incompleteness and crowding effects as estimated with the artificial star tests.
For a better assessment of the uncertainties involved in the SFH derivation, we will apply three different and independent approaches of the synthetic CMD method:  \citep[see e.g.,][]{tosi91,cole07,cignoni09}. 







\acknowledgments

We thank Gisella Clementini and Luigi (Rolly) Bedin for their useful suggestions, and Livia Origlia for providing unpublished photometric conversion tables. Support for Program GO-10396 was provided by NASA through a grant from the Space Telescope Science Institute, which is operated by the Association of Universities for Research in Astronomy, Incorporated, under NASA contract NAS5-26555. JSG thanks NASA/STScI for their support of this research, which also benefited from funding supplied by PRIN-INAF-2005, ASI-INAF I/016/07/0 and Swiss National Science Foundation grant 200020-113697.

\clearpage


\clearpage

\begin{deluxetable}{ccllcll}
\tabletypesize{\scriptsize}

\tablecaption{Journal of Observations\label{t:obs}}
\tablewidth{0pt}
\tablehead{
\colhead{SMC field} & \colhead{Image Name} & \colhead{R.A. (J2000)} & \colhead{Decl.
(J2000.)} & \colhead{Date of Obs.} &
\colhead{Exp. Time (s)} & \colhead{Filter}
}
\startdata
SFH1  & J96110010 & 00$^h$51$^m$39.31$^s$ & -73$\degr$07$\arcmin$00.9$\arcsec$ & 2006 Jan 17 & $2\times 1.0$    & F814W \\
Bar (SW)& J96110020 & 00$^h$51$^m$39.36$^s$ & -73$\degr$07$\arcmin$02.8$\arcsec$ & 2006 Jan 17 & $3\times 504.0$ & F814W \\
       & J96110030 & 00$^h$51$^m$39.31$^s$ & -73$\degr$07$\arcmin$00.9$\arcsec$ & 2006 Jan 17 & $2\times 1.0$    & F555W \\
       & J96110040 & 00$^h$51$^m$39.36$^s$ & -73$\degr$07$\arcmin$02.8$\arcsec$ & 2006 Jan 17 & $3\times 539$.0 & F555W \\
       & J96110DTQ & 00$^h$51$^m$40.00$^s$ & -73$\degr$07$\arcmin$01.0$\arcsec$ & 2006 Jan 17 & $1\times 48.0$   & F814W \\
       & J96110E5Q & 00$^h$51$^m$40.00$^s$ & -73$\degr$07$\arcmin$01.0$\arcsec$ & 2006 Jan 17 & $1\times 50.0$   & F555W \\
SFH10 & J96112010 & 01$^h$08$^m$39.00$^s$ & -72$\degr$58$\arcmin$45.0$\arcsec$ & 2006 Jan 17 & $2\times 1.0$    & F814W \\
Wing (SE)& J96112020 & 01$^h$08$^m$39.18$^s$ & -72$\degr$58$\arcmin$46.9$\arcsec$ & 2006 Jan 17 & $3\times 504.0$ & F814W \\
       & J96112030 & 01$^h$08$^m$39.12$^s$ & -72$\degr$58$\arcmin$45.0$\arcsec$ & 2006 Jan 17 & $2\times 1.0$    & F555W \\
       & J96112040 & 01$^h$08$^m$39.18$^s$ & -72$\degr$58$\arcmin$46.9$\arcsec$ & 2006 Jan 17 & $3\times 539.0$ & F555W \\
       & J96112FJQ & 01$^h$08$^m$39.80$^s$ & -72$\degr$58$\arcmin$45.0$\arcsec$ & 2006 Jan 17 & $1\times 48.0$   & F814W \\
       & J96112FVQ & 01$^h$08$^m$39.80$^s$ & -27$\degr$58$\arcmin$45.0$\arcsec$ & 2006 Jan 17 & $1\times 50.0$   & F555W \\
SFH4  & J96111010 & 01$^h$00$^m$02.63$^s$ & -72$\degr$32$\arcmin$06.5$\arcsec$ & 2005 Nov 23 & $2\times 1.0$    & F814W \\ 	  
Bar (center)& J96111020 & 01$^h$00$^m$03.01$^s$ & -72$\degr$32$\arcmin$07.3$\arcsec$ & 2005 Nov 23 & $3\times 504.0$ & F814W \\
       & J96111030 & 01$^h$00$^m$02.63$^s$ & -72$\degr$32$\arcmin$06.5$\arcsec$ & 2005 Nov 23 & $2\times 1.0$    & F555W \\
       & J96111040 & 01$^h$00$^m$03.01$^s$ & -72$\degr$32$\arcmin$07.3$\arcsec$ & 2005 NOV 23 & $3\times 539.0$ & F555W \\
       & J96111LOQ & 01$^h$00$^m$03.00$^s$ & -72$\degr$32$\arcmin$04.0$\arcsec$ & 2005 Nov 23 & $1\times 48.0$   & F814W \\
       & J96111M0Q & 01$^h$00$^m$03.00$^s$ & -72$\degr$32$\arcmin$04.0$\arcsec$ & 2005 Nov 23 & $1\times 50.0$   & F555W \\
SFH5  & J96116010 & 01$^h$00$^m$24.31$^s$ & -73$\degr$11$\arcmin$09.9$\arcsec$ & 2006 Jan 18 & $2\times 10.0$   & F814W \\
Bar   & J96116020 & 01$^h$00$^m$24.54$^s$ & -73$\degr$11$\arcmin$07.8$\arcsec$ & 2006 Jan 18 & $4\times 474.0$ & F814W \\
       & J96116030 & 01$^h$00$^m$24.31$^s$ & -73$\degr$11$\arcmin$09.9$\arcsec$ & 2006 Jan 18 & $2\times 20.0$   & F555W \\
       & J96116040 & 01$^h$00$^m$24.54$^s$ & -73$\degr$11$\arcmin$07.8$\arcsec$ & 2006 Jan 18 & $4\times 496.0$ & F555W \\
SFH8  & J96114010 & 00$^h$41$^m$47.58$^s$ & -71$\degr$03$\arcmin$47.2$\arcsec$ & 2005 Nov 29 & $2\times 10.0$   & F814W \\
Halo (N)       & J96114020 & 00$^h$41$^m$47.38$^s$ & -71$\degr$03$\arcmin$45.0$\arcsec$ & 2005 Nov 29 & $4\times 474.0$ & F814W \\
       & J96114030 & 00$^h$41$^m$47.58$^s$ & -71$\degr$03$\arcmin$47.2$\arcsec$ & 2005 Nov 29 & $2\times 20.0$   & F555W \\
       & J96114040 & 00$^h$41$^m$47.38$^s$ & -71$\degr$03$\arcmin$45.0$\arcsec$ & 2005 Nov 29 & $4\times 496.0$ & F555W \\
SFH9 & J96115010 & 01$^h$21$^m$52.68$^s$ & -73$\degr$18$\arcmin$48.7$\arcsec$ & 2005 Nov 21 & $2\times 10.0$   & F814W \\
Wing/Bridge& J96115020 & 01$^h$21$^m$52.33$^s$ & -73$\degr$18$\arcmin$46.8$\arcsec$ & 2005 Nov 21 & $4\times 474.0$ & F814W \\
       & J96115030 & 01$^h$21$^m$52.68$^s$ & -73$\degr$18$\arcmin$48.7$\arcsec$ & 2005 Nov 21 & $2\times 20.0$   & F555W \\
       & J96115040 & 01$^h$21$^m$52.33$^s$ & -73$\degr$18$\arcmin$46.8$\arcsec$ & 2005 Nov 21 & $4\times 496.0$ & F555W \\
 \enddata
\end{deluxetable}

\begin{deluxetable}{lllll}
\tabletypesize{\scriptsize}

\tablecaption{Properties of the fields\label{t:fields}}
\tablewidth{0pt}
\tablehead{
\colhead{SMC field} & \colhead{E(B-V)} & \colhead{RC mag} & \colhead{(m-M)$_0$} & \colhead{RC color}}
\startdata

SFH1 & 0.08 & 18.45$\pm$0.45 & 18.89 & 1.00$\pm$0.06\\
SFH4 & 0.07 & 18.55$\pm$0.27 & 18.97 & 1.03$\pm$0.05\\
SFH5 & 0.06 & 18.56$\pm$0.28 & 18.96 & 0.99$\pm$0.04\\ 
SFH10 & 0.06 & 18.49$\pm$0.31 & 18.89 & 1.00$\pm$0.04\\
SFH9 & 0.07 & 18.48$\pm$0.28 & 18.90 & $\sim 0.98$\\
SFH8 & 0.07 & 18.56$\pm$0.27 & 18.95 & $\sim 0.95$\\

\enddata
\end{deluxetable}


\end{document}